\documentclass[aps,prd,reprint,groupedaddress,superscriptaddress,preprintnumbers,floatfix,a4paper,nofootinbib]{revtex4-1}
\usepackage{color}
\usepackage{amsmath}
\usepackage{amssymb}
\usepackage{dcolumn}  
\usepackage{slashed}
\usepackage{isotope}
\usepackage{graphicx}
\usepackage{mathtools}
\usepackage[utf8]{inputenc}
\usepackage{multirow}
\usepackage{hyperref}
\hypersetup{colorlinks, linkcolor = [rgb]{0,0.0,0.75}, citecolor = [rgb]{0,0.0,0.75}, urlcolor = [rgb]{0,0.0,0.75}}

\makeatletter
\g@addto@macro\bfseries{\boldmath}
\makeatother

\newcommand{\beq}{\begin{equation}}
\newcommand{\eeq}{\end{equation}}

\newcommand{\bea}{\begin{eqnarray}}
\newcommand{\eea}{\end{eqnarray}}

\newcommand{\bee}{\begin{enumerate}}
\newcommand{\eee}{\end{enumerate}}

\newcommand{\bef}{\begin{figure}}
\newcommand{\eef}{\end{figure}}

\newcommand{\bei}{\begin{itemize}}
\newcommand{\eei}{\end{itemize}}

\newcommand{\ripm}{\text{RI$'$\kern-0.06667em/MOM}}

\if@mathematic
\def\vec#1{\ensuremath{\mathchoice
    {\mbox{\boldmath$\displaystyle\mathbf{#1}$}}
    {\mbox{\boldmath$\textstyle\mathbf{#1}$}}
    {\mbox{\boldmath$\scriptstyle\mathbf{#1}$}}
    {\mbox{\boldmath$\scriptscriptstyle\mathbf{#1}$}}}}
\else
\def\vec#1{\ensuremath{\mathchoice
    {\mbox{\boldmath$\displaystyle#1$}}
    {\mbox{\boldmath$\textstyle#1$}}
    {\mbox{\boldmath$\scriptstyle#1$}}
    {\mbox{\boldmath$\scriptscriptstyle#1$}}}}
\fi

\newcommand{\eVdist}{\kern-0.06667em}

\newcommand{\eq}[1]{Eq.\,({\ref{#1}})}

\begin{document}
\allowdisplaybreaks

\preprint{CERN-TH-2018-098}
\preprint{DESY 18-066}
\preprint{HIM-2018-02}
\preprint{MITP/18-030}
\preprint{TIFR/TH/18-12}

\title{Lattice QCD study of the $H$\,dibaryon using hexaquark and two-baryon interpolators}



\author{A.~Francis} 
\affiliation{Theoretical Physics Department, CERN, CH-1211 Geneva 23, Switzerland}

\author{J.\,R.~Green}
\affiliation{NIC, Deutsches Elektronen-Synchrotron, D-15738 Zeuthen, Germany}

\author{P.\,M.~Junnarkar}
\affiliation{Tata Institute of Fundamental Research (TIFR), 1 Homi Bhabha Road, Mumbai 400005. India.}

\author{Ch.~Miao}
\affiliation{PRISMA Cluster of Excellence and Institut f\"ur Kernphysik, University of Mainz, Becher Weg 45, D-55099 Mainz, Germany}
\affiliation{Helmholtz Institute Mainz, University of Mainz, D-55099 Mainz, Germany}

\author{T.\,D.~Rae}
\affiliation{PRISMA Cluster of Excellence and Institut f\"ur Kernphysik, University of Mainz, Becher Weg 45, D-55099 Mainz, Germany}

\author{H.~Wittig}
\affiliation{PRISMA Cluster of Excellence and Institut f\"ur Kernphysik, University of Mainz, Becher Weg 45, D-55099 Mainz, Germany}
\affiliation{Helmholtz Institute Mainz, University of Mainz, D-55099 Mainz, Germany}


\date{\today}


\begin{abstract}
We present a lattice QCD spectroscopy study in the isospin singlet,
strangeness $-2$ sectors relevant for the conjectured $H$~dibaryon. We
employ both local and bilocal interpolating operators to isolate the
ground state in the rest frame and in moving frames. Calculations are
performed using two flavors of $O(a)$-improved Wilson fermions and a
quenched strange quark. Our initial point-source method for
constructing correlators does not allow for bilocal operators at the
source; nevertheless, results from using these operators at the sink
indicate that they provide an improved overlap onto the ground state
in comparison with the local operators. We also present results, in
the rest frame, using a second method based on distillation to compute
a hermitian matrix of correlators with bilocal operators at both the
source and the sink. This method yields a much more precise and
reliable determination of the ground-state energy. In the flavor-SU(3)
symmetric case, we apply Lüscher's finite-volume quantization
condition to the rest-frame and moving-frame energy levels to
determine the $S$-wave scattering phase shift, near and below the
two-particle threshold. For a pion mass of 960\,MeV, we find that
there exists a bound $H$~dibaryon with binding energy
${\Delta}E=(19\pm10)$\,MeV. In the 27-plet (dineutron) sector, the
finite-volume analysis suggests that the existence of a bound state is
unlikely.
\end{abstract}



\maketitle


\section{Introduction}

The strong force between quarks and gluons produces a rich spectrum of
bound states and resonances, the color-neutral hadrons. Most of these
can be described by constituent quark models as either quark-antiquark
mesons or three-quark baryons. The existence of exotic hadrons, which
cannot be described as such, is an active field of inquiry. Over the
past several years, the so-called ``X, Y, Z'' mesons have been studied
intensively, both theoretically and
experimentally~\cite{Lebed:2016hpi}, and in recent years pentaquark
baryons have also gained attention~\cite{Aaij:2015tga}.

Nearly four decades ago, using the MIT bag model, Jaffe predicted a
deeply bound dibaryon with quark content $uuddss$ that is a scalar and
a flavor singlet, the $H$\,dibaryon~\cite{Jaffe:1976yi}. In contrast
with the only known stable dibaryon, the deuteron, which can be well
described as a loosely-bound proton-neutron state and is bound by just
2.2~MeV, the bag model predicted the $H$~dibaryon as an exotic
hexaquark state where all six quarks are in S-wave in the same
hadronic bag, bound by about 80~MeV below the $\Lambda\Lambda$
threshold.

Experimental evidence disfavors such a large binding energy. The
strongest constraint is the ``Nagara'' event provided by the E373
experiment at KEK~\cite{Takahashi:2001nm}, which found a
$\isotope[6][\Lambda\Lambda]{He}$ double-hypernucleus with
$\Lambda\Lambda$ binding energy $B_{\Lambda\Lambda}=6.91\pm
0.16$~MeV~\cite{Nakazawa:2010zza} that decayed weakly. A deeply bound
$H$~dibaryon would enable the strong decay
$\isotope[6][\Lambda\Lambda]{He} \to \isotope[4]{He} + H$; its absence
implies $m_H > 2m_\Lambda - B_{\Lambda\Lambda}$. There was also no
indication of an $H$~dibaryon from a high-statistics study of upsilon
decays at Belle~\cite{Kim:2013vym}.

The first lattice QCD study of the $H$~dibaryon was performed more
than thirty years ago~\cite{Mackenzie:1985vv}, using a quenched
ensemble with lattice size $6^2\times 12\times 18$. Quenched studies
--- which all used local interpolating operators with six quarks at
the same point in keeping with the bag model picture, together with
standard lattice spectroscopy techniques --- produced inconclusive
results: while some found a bound
state~\cite{Iwasaki:1987db,Luo:2007zzb,Luo:2011ar}, others did
not~\cite{Mackenzie:1985vv,Pochinsky:1998zi,Wetzorke:1999rt,Wetzorke:2002mx}.
Early studies of the $H$~dibaryon using lattice QCD are summarized in
Ref.~\cite{Beane:2011zpa}.

Aside from the present work\footnote{Exploratory studies and
  preliminary results were previously reported in
  Refs.~\cite{Francis:2013lva,Green:2014dea,Junnarkar:2015jyf}.},
calculations with dynamical fermions have been performed by two
collaborations, both of which reported a bound $H$~dibaryon at
heavier-than-physical quark masses. The NPLQCD collaboration performed
lattice spectroscopy calculations using a setup based on clover
fermions with local hexa\-quark operators at the source and bilocal
two-baryon operators at the sink. First results were obtained on
anisotropic ensembles with $N_f=2+1$ dynamical
fermions~\cite{Beane:2010hg,Beane:2011zpa,Beane:2011iw}, followed by
isotropic ensembles with three mass-degenerate ($N_f=3$)
quarks~\cite{Beane:2012vq}. An alternative approach, employed by the
HAL~QCD collaboration, is based on determining baryon-baryon
potentials from Nambu-Bethe-Salpeter wave functions computed on the
lattice, followed by solving the Schr\"odinger equation to study
baryon-baryon scattering and bound states. This was done on ensembles
with $N_f=3$ clover fermions for a range of quark
masses~\cite{Inoue:2010hs,Inoue:2010es,Inoue:2011ai}. Although these
two sets of calculations agreed on the presence of a bound state, they
disagreed significantly on the binding energy: in the $N_f=3$ case
with pseudoscalar meson mass near 800~MeV, the value reported by
NPLQCD was $74.6\pm4.7$\,MeV, whereas HAL~QCD reported
$37.8\pm5.1$\,MeV. Recently, HAL~QCD have published a $N_f=2+1$ study
of coupled channel ($\Lambda\Lambda$ and $N\Xi$) baryon-baryon
interactions with near-physical quark masses, which claims that the
$H$~dibaryon may be a $\Lambda\Lambda$ resonance just below the $N\Xi$
threshold~\cite{Sasaki:2016gpc,Sasaki:2018mzh}.

Given that there are conflicting results for the binding energy of the
$H$~dibaryon, we have started a new initiative which may help to
resolve the issue. As a first step we present results from a study in
two-flavor QCD, i.e.\ with a mass-degenerate doublet of dynamical $u$
and $d$ quarks. The mass of the (quenched) strange quark is either
tuned such that $m_s=m_d=m_u$ or set to a heavier value, implying that
the SU(3) flavor symmetry is broken. Clearly, SU(3) symmetry is
significantly broken at the physical point
\cite{Haidenbauer:2011ah,Haidenbauer:2011za}, which allows the three
flavor multiplets, i.e.\ the singlet, octet and 27-plet to
couple. Therefore, it is advantageous to study the octet and 27-plet
even in the case of exact SU(3) symmetry. Furthermore, the 27-plet
contains the two-nucleon $I=1$ sector which has a possible dineutron
bound state. The nucleon-nucleon sector has been studied extensively
in experiment and may serve as a benchmark for lattice calculations.

Our work is mainly focused on the methodology of determining the
spectrum and the binding energy via the computation of correlation
matrices and their
diagonalization~\cite{Michael:1985ne,Luscher:1990ck,Blossier:2009kd}. In
order to allow for a direct comparison with the results from older
quenched studies and from NPLQCD we have chosen a similar setup. As we
will describe in more detail in the following sections, we have used
point sources to compute correlator matrices with local interpolating
operators at the source and both local and bilocal interpolators at
the sink. In addition, we report initial results from a follow-up
study in which we, for the first time, applied the distillation
method~\cite{Peardon:2009gh} to the two-baryon sector. This allowed us
to compute a correlator matrix using operators made from products of
two spatially displaced, momentum-projected baryon interpolators at
both the source and the sink. We shall see that this hermitian setup
leads to a more robust and precise identification of the spectrum.

Since the strange quark is quenched in our calculation, one may think
that any observed deviation from the findings of
Refs.\,\cite{Beane:2010hg, Beane:2011zpa, Beane:2011iw, Beane:2012vq,
  Inoue:2010hs, Inoue:2010es, Inoue:2011ai} should be attributed to
the different treatment of the quark sea. However, the FLAG
report~\cite{Aoki:2016frl} provides ample evidence that observables
computed with $N_{\text{f}}=2$ or $N_{\text{f}}=2+1$ dynamical quarks
differ at the percent level at most.

This paper is organized as follows. Our methodology is described in
Section~\ref{setup}: this includes the interpolating operators and our
approach for analyzing correlator matrices. We show our determination
of the energy levels using point-source methods in
Section~\ref{sec:analysis_point} and using distillation in
Section~\ref{sec:analysis_distillation}. In
Section~\ref{sec:finite_volume}, we apply L\"uscher's finite-volume
quantization condition to determine scattering phase shifts at the
SU(3)-symmetric point, and identify the presence of a bound
$H$~dibaryon. Finally, our conclusions are presented in
Section~\ref{sec:conclusions}.

\section{\label{setup} Lattice calculation and setup}

\subsection{Simulation details}
\label{sec:numerical_setup}

Our study has been performed on a set of ensembles with two
mass-degenerate dynamical flavors of O($a$)-improved Wilson quarks
\cite{Sheikholeslami:1985ij} and the Wilson plaquette action, which
were generated as part of the CLS (Coordinated Lattice Simulations)
initiative, using the deflation-accelerated DD-HMC
\cite{Luscher:2005rx,Luscher:2007es} and MP-HMC
\cite{Marinkovic:2010eg} algorithms. The improvement coefficient
$c_{\rm sw}$ multiplying the Sheikholeslami-Wohlert term was tuned
according to the non-perturbative determination of
Ref.~\cite{Jansen:1998mx}. An overview of the ensembles can be found
in Table~\ref{tab:lat_par1}. All our calculations were performed in
the SU(3)-flavor symmetric limit, with the exception of ensemble E5
for which the valence strange quark mass was tuned so that the
combination $(2m_K^2-m_\pi^2)/m_\Omega^2$ takes its physical
value. The corresponding values of $m_\pi$ and $m_K$ are provided in
the table. The values of the lattice spacing in physical units were
determined using the kaon decay constant \cite{Fritzsch:2012wq}.

Quark propagators were computed using the Schwarz alternating
procedure (SAP) domain-decomposed, deflated generalized conjugate
residual (GCR) solver of the DD-HMC package \cite{Luscher:2005rx} with
smeared point sources on a grid of source positions that was randomly
displaced on each gauge configuration. For the distillation
calculation, we used a similar solver in OpenQCD~\cite{OpenQCD},
computed low modes of the spatial Laplacian using
PRIMME~\cite{PRIMME}, and contracted them to form ``perambulators''
and mode triplets using QDP++~\cite{Edwards:2004sx}.  Baryon and
multi-baryon correlators computed in lattice QCD suffer from a severe
signal-to-noise problem, since the noise grows with a rate
proportional to $\exp\{({m_{\rm B}-3/2 m_\pi)t}\}$ per baryon, where
$m_{\rm B}$ denotes the baryon mass. This makes it difficult to
identify a ``window'' in which the asymptotic behavior has been
reached while the signal is not yet lost in the statistical noise. In
order to allow for a significant increase in statistics while keeping
the numerical effort at a manageable level, we have employed the
method of all-mode-averaging (AMA) \cite{Blum:2012uh}. This entails
computing a high number of samples with lower-precision propagator
solves, followed by applying a bias correction using a relatively
small number of high-precision solves. We used this for the
calculation with point-source propagators, but obtained only modest
cost savings due to our use of a highly efficient solver.

\begingroup
\renewcommand*{\arraystretch}{1.25}
\begin{table*}[t!]
\centering
\begin{tabular}{c|cccc| cc}
\hline\hline
& \multicolumn{4}{c|}{point-to-all} & \multicolumn{2}{c}{timeslice-to-all}\\
Label & N1 & E5 & E1 & A1 & E5 & E1 \\ \hline
Size & $\:48^3\times96\:$ & $\:32^3\times64\:$ & $\:32^3\times64\:$ & $\:32^3\times64\:$ & $\:32^3\times64\:$ & $\:32^3\times64\:$   \\
$\beta$ & 5.5 & 5.3 & 5.3 & 5.2 & 5.3 & 5.3 \\
$a\;\left[\text{fm}\right]$ & 0.0486(6) & 0.0658(10) & 0.0658(10) & 0.0755(11) & 0.0658(10) & 0.0658(10) \\
$m_\pi\;\left[\text{MeV}\right]$ & 858 & 436 & 960 & 744 & 436 & 960 \\
$m_K\;\left[\text{MeV}\right]$   & 858 & 648 & 960 & 744 & 648 & 960 \\
$L\;\left[\text{fm}\right]$ & 2.33 & 2.11 & 2.11 & 2.42 & 2.11 & 2.11  \\
$m_\pi L$ & 10.0 & 4.7 & 10.2 & 9.9 & 4.7 & 10.2  \\
$N_{\text{conf}}$ & 100 & 1990 & 168 & 286 & 2000 & 168 \\
$N_{\text{src}}|N_{\text{tsrc}}$ & 128 & 32 & 128 & 128 & 4 & 8  \\
$N_{\text{meas}}$ & 25600 & 127360 & 43008 & 73216 & 16000 & 2688 \\
\hline\hline
\end{tabular}
\caption{Overview of ensemble parameters used in this study. For
  ensembles N1, E1 and A1 all quark masses were tuned to realize the
  SU(3)-flavor symmetric case, while for E5 the quark masses are
  non-degenerate. $N_{\text{src}}$ denotes the number of sources in
  the calculation of point-to-all propagators using the AMA method,
  while $N_{\text{tsrc}}$ is the number of timeslices used to compute
  timeslice-to-all propagators with the distillation method. For every
  configuration and source position we have computed the correlators
  in the forward and backward directions, resulting in two independent
  measurements.}
\label{tab:lat_par1}
\end{table*}
\endgroup

\subsection{Interpolating operators \label{sec:interpolators}} 

Accurate determinations of the spectrum in the $H$~dibaryon channel
require a set of efficient interpolating operators whose projection
properties onto the ground state may also give qualitative insights
into the nature of the $H$~dibaryon. For instance, the local operators
defined in Equation~\eqref{eq:hexaquark} below resemble more closely
Jaffe's original interpretation of the $H$~dibaryon as a deeply bound
state of six quarks forming a color singlet. By contrast, bilocal
two-baryon operators (see Equation~\eqref{eq:2baryonop}) may be more
appropriate to describe loosely-bound states such as the
deuteron. While the two types of operators are defined according to a
qualitative physical picture that is suggestive of the nature of a
given state, this is not a rigorous way to study its properties.

The generic form of a lattice QCD correlation function is given by
\begin{equation}
   C_{ij}(\vec{P},\tau) = \left\langle
   \mathcal{O}_i({\vec{P}},t)
   \mathcal{O}_j({\vec{P}},t')^\dagger \right\rangle,\quad
   \tau=t-t^\prime,
\label{eq:corrmatrix}
\end{equation}
where the interpolating operator $\mathcal{O}_i$ carries the quantum
numbers of the continuum state under study, and it is understood that
$\mathcal{O}_i$ has been projected onto spatial momentum $\vec{P}$.
When constructing interpolators in the $H$~dibaryon channel, one can
think of two generic configurations. Jaffe's original analysis was
based on a compact color-singlet comprising six quarks, which gives
rise to a hexaquark operator composed of flavors $uuddss$. The
alternative possibility is the product of two individual
color-singlets at different positions, i.e.\ a two-baryon operator.

The starting point for the construction of local hexaquark operators
is the object
\begin{align}\label{eq:hexaquark}
[rstuvw] = \epsilon_{ijk} &\epsilon_{lmn} \Big( s^i C\gamma_5 P_+ t^j
\Big) \\ \nonumber
& \times\Big( v^l C\gamma_5 P_+ w^m \Big) \Big( r^k C\gamma_5 P_+ u^n \Big) ({\vec{x}}, t)\,,
\end{align}
where $r, s,\ldots,w$ denote generic quark flavors, and
$P_+=(1+\gamma_0)/2$ projects the quark fields to positive parity. One
can form two operators that transform under the
singlet\,\cite{Donoghue:1986zd,Golowich:1992zw,Wetzorke:1999rt} and
27-plet irreducible representations of flavor SU(3):
\begin{align}
H_\mathbf{1} &= \frac{1}{48}\Big( [sudsud] - [udusds] - [dudsus]  \Big), \\
H_\mathbf{27}&= \frac{1}{48\sqrt{3}}\Big( 3[sudsud] + [udusds] + [dudsus]  \Big)\,.
\end{align}
The continuum quantum numbers of these operators are $S=-2$,
$I(J^{P})=0(0^{+})$, as required by the original bag-model
proposal. Equation~\eqref{eq:hexaquark} is the product of two
single-baryon operators with quark content $rst$ and $uvw$ at the same
point; one could then interpret $H_\mathbf{1}$ and $H_\mathbf{27}$ as
linear combinations of $\Lambda\Lambda$ and
$p\Xi^{-}+n\Xi^{0}$. However, as argued in
Ref.~\cite{Golowich:1992zw}, this is not meaningful because
antisymmetrization of the six quarks implies that the operators are
also equal to linear combinations of $\Lambda\Lambda$ and
$\Sigma\Sigma$. Similarly, $H_\mathbf{1}$ could also be written as the
product of two color-octet triquarks~\cite{Donoghue:1986zd} or of
three diquarks~\cite{Jaffe:2004ph}. We stress that it is the unique
SU(3)-singlet scalar operator made from six positive-parity-projected
quarks at the same point. We also note that an octet state cannot be
represented in terms of this simplest class of hexaquark
operators. Finally, we project onto the momentum of each lattice
frame:
\begin{equation}
H_{\{\mathbf{1,27}\}}(\vec{P},t) = \sum_{\vec{x}} e^{-i\vec{P\cdot x}}
H_{\{\mathbf{1,27}\}}(\vec{x},t).
\end{equation}
From now on, we will refer to this type of interpolator as a local
hexaquark operator.

An alternative configuration in the $H$~dibaryon
channel is described by the product of two spatially displaced,
momentum-projected single baryon operators \cite{Inoue:2010hs}:
\begin{align}\label{eq:2baryonop}
  &\big(BB\big)_m({\vec{P}},t) =
  \sum_{\vec{p_1},\vec{p_2}} f_m(\vec{p_1},\vec{p_2})
  \\ \nonumber
  &\times    \sum_{{\vec{x}}}
  e^{-i{\vec{p_1\cdot x}}} B_{1\alpha}({\vec{x}},t) 
 (C\gamma_5P_+)_{\alpha\beta} \sum_{{\vec{y}}}
  e^{-i{\vec{p_2\cdot y}}} B_{2\beta}({\vec{y}},t)\,, 
\end{align}
where
\begin{equation}
  B_\alpha=[rst]_\alpha = \epsilon_{ijk} \big( s^i C\gamma_5 P_+ t^j
  \big) r^k_\alpha\,,
\end{equation}
and the individual baryons have been projected onto spatial momenta
$\vec{p}_1$ and $\vec{p}_2$ with total momentum
$\vec{P}=\vec{p}_1+\vec{p}_2$. The index~$m$ labels a particular
configuration of momenta $\vec{p}_1$ and $\vec{p}_2$. In the rest
frame, the momentum combinations have a particularly simple
construction:
\begin{equation}
  f_m(\vec{p}_1,\vec{p}_2) = \begin{cases}
    1 & \vec{p}_1=-\vec{p}_2 \text{ and } \vec{p}_1^2=m\left({2\pi}/{L}\right)^2 \\
    0 & \text{otherwise}
    \end{cases}.
\end{equation}
In the remainder of this paper, we refer to the object defined in
Equation~\eqref{eq:2baryonop} as a bilocal two-baryon operator.

We form dibaryon operators from combinations of octet baryons with
$S=-2$, $I=0$, i.e.
\begin{align}
\big(\Lambda \Lambda\big) &= \frac{1}{12}
  [sud][sud]~~, \label{eq:LL}\\ 
\begin{split} \label{eq:NXS}
\big(N \Xi\big)_S &= \frac{1}{36} \Big( [uud][ssd] - [dud][ssu]\\
&\qquad + [ssd][uud] - [ssu][dud] \Big)~~,
\end{split}\\
\begin{split} \label{eq:SS}
\big(\Sigma \Sigma\big) &= \frac{1}{36\sqrt{3}} \Big( 2[uus][dds] -
  [dus][uds] \\
&\qquad - [dus][dus] - [uds][dus] \\
&\qquad- [uds][uds] + 2[dds][uus]  \Big)~,
\end{split}
\end{align}
where the subscript $S$ on $(N\Xi)$ denotes the flavor-symmetric
combination.\footnote{While one can construct a flavor-antisymmetric
  combination $(N\Xi)_A$, one finds that such a state is excluded in
  infinite volume, because the overall antisymmetry prevents it from
  having $J^P=0^+$. A more technical reason for ignoring it is the
  fact that the state belongs to the flavor octet which we find
  difficult to resolve even for the flavor-symmetric combination.}

Using the rotation matrices listed in Appendix~B of
Ref.\,\cite{Inoue:2010hs} we form the appropriate linear combinations
of $(\Lambda\Lambda)$, $(N\Xi)_S$ and $(\Sigma\Sigma)$ that correspond
to different flavor multiplets, i.e.\ the singlet, octet and
27-plet. The projected operators are then called
$BB_{\{\mathbf{1,8,27}\},m}$. 

In the interpolating operators defined above, we use smeared quark
fields; the smearing helps to increase the coupling of an operator to
the low-lying states. For the point-source calculation, we used
Wuppertal smearing~\cite{Gusken:1989qx}, with the hopping term
constructed using spatially APE-smeared~\cite{Albanese:1987ds} gauge
links. To increase the size of our operator basis, we used two
different smearing widths: ``narrow'' (70 steps, denoted $N$) and
``medium'' (140 steps, denoted $M$). The distillation approach makes
use of Laplacian-Heaviside (LapH) smearing~\cite{Peardon:2009gh}, in
which the quark fields are smeared by projecting them onto the
low-lying modes of the spatial gauge-covariant Laplacian, itself
constructed using stout-smeared~\cite{Morningstar:2003gk} gauge links.
We used 56 modes in all cases, and label this type of smearing
as~$L$. We include the smearing as part of the label for each
interpolating operator, yielding names such as $H_{\mathbf{1},N}$ or
$BB_{\mathbf{27},L,0}$.

\subsection{Operator basis and correlation
  matrices \label{sec:matrices}} 

The determination of hadronic energy levels in lattice QCD usually
proceeds by computing a correlation matrix $C_{ij}(t)$ for the chosen
basis of interpolating operators [see Equation~\eqref{eq:corrmatrix}]
and solving a generalized eigenvalue problem (GEVP)
\cite{Michael:1985ne,Luscher:1990ck,Blossier:2009kd}.

In the following the correlation matrices
$C_{ij}(t)=C_{ij}(t,{\vec{P}})$ are evaluated at rest, i.e.\  with
total momentum ${\vec{P}}=0$, and in moving frames,
i.e.\ ${\vec{P}}^2>0$.  We included two-baryon operators with
individual momenta $\vec{p}_{1,2}=2\pi\vec{d}_{1,2}/L$ and
$\vec{d}^2=0, 1, 2, 3$ in the calculation. The components were chosen
such as to realize a total momentum $\vec{P}=\vec{p}_1+\vec{p}_2\equiv
2\pi\vec{D}/L$ with $\vec{D}^2=0, 1, 2, 3$. For each $\vec{D}^2$, we
average over all equivalent frames that are related by a lattice
rotation.

\begin{figure}
  \centering
  \includegraphics[width=0.3\columnwidth]{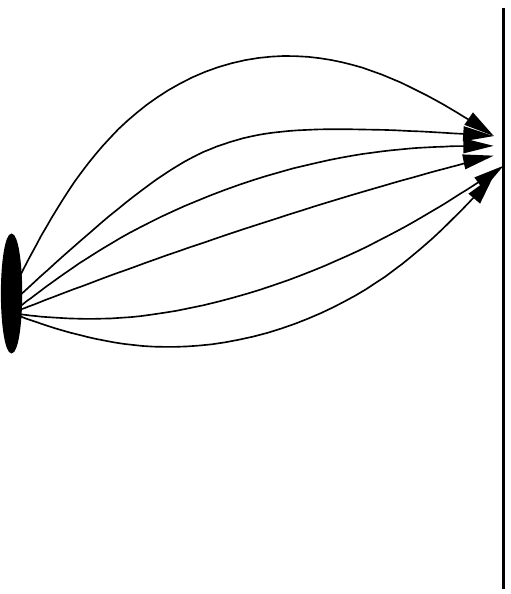}
  \hspace{0.2\columnwidth}
  \includegraphics[width=0.3\columnwidth]{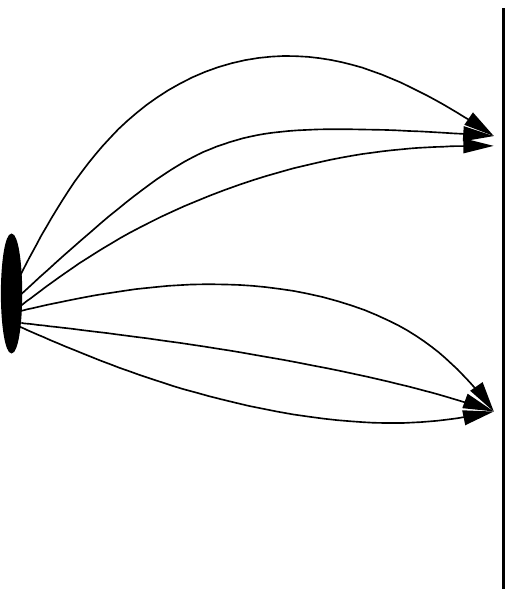}\\
  $\langle H(t) H^\dagger(0)\rangle$
  \hspace{0.3\columnwidth}
  $\langle BB(t) H^\dagger(0) \rangle$
  \caption{Quark lines for correlators using point sources. The
    vertical lines indicate that either the single or the two
    different end points of the quark lines are summed over the sink
    timeslice.}
  \label{fig:quark_lines_point}
\end{figure}

\begin{figure}
  \centering
  \includegraphics[width=0.3\columnwidth]{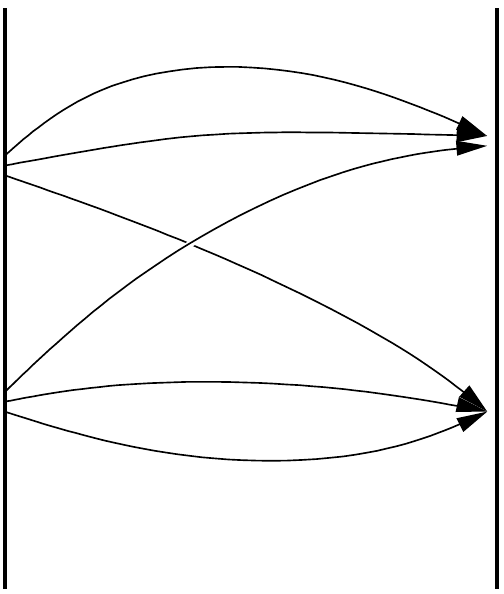}\\
  $\langle BB(t) BB^\dagger(0) \rangle$
  \caption{Quark lines showing an example contraction for correlators
    using distillation. The vertical lines indicate that the two start
    and two end points for the quark lines are summed over the source
    and sink timeslice, respectively.}
  \label{fig:quark_lines_distillation}
\end{figure}

Performing the Wick contractions of correlators involving bilocal
two-baryon operators of the type in \eq{eq:2baryonop} results in
diagrams that contain quark lines that start or end at two distinct
spatial points within a timeslice, This has important consequences for
computing correlation matrices, since point-to-all propagators do not
allow for the computation of such diagrams whenever a two-baryon
operator is placed at the source. We therefore opted for an asymmetric
setup in which only hexaquark operators were put at the source, while
at the sink both hexaquark and two-baryon operators were used; this is
illustrated in Figure~\ref{fig:quark_lines_point}. In this setting,
the familiar GEVP or, more generally, the diagonalization procedure
must be modified in order to allow for a non-hermitian correlator
matrix. We select subsets of $N_{\textrm{op}}$ source operators and
$N_{\textrm{op}}$ sink operators and form the corresponding square
correlator matrix $\mathbf{C}(t)$.  We then perform the following
steps, starting from this $N_{\textrm{op}}\times N_{\textrm{op}}$
matrix:
\begin{enumerate}
\item Determine the right and left eigenvectors of $\mathbf{C}(t)$ by
  solving 
  \begin{align}
    \mathbf{C}(t_1)v_n(t_1,t_0) &=
    \lambda_n(t_1,t_0)\,\mathbf{C}(t_0)v_n(t_1,t_0), \\
    w_n^\dagger(t_1,t_0)\mathbf{C}(t_1) &=
    \lambda_n(t_1,t_0)\,w_n^\dagger(t_1,t_0)\mathbf{C}(t_0),
  \end{align}
  for $n=1,\ldots,N_{\textrm{op}}$, where $t_0$ and $t_1$ denote
  fixed timeslices in the region where the lowest $N_{\textrm{op}}$
  states are expected to dominate.
\item Compute the (approximately) diagonal matrix
  $\mathbf{\Lambda}(t)$ whose elements are given by\footnote{Note that
  $\mathbf{\Lambda}(t)$ is exactly diagonal for $t=t_0$ and $t=t_1$.}
  \begin{equation}
    \Lambda_{nm}(t) = w_n^\dagger \mathbf{C}(t) v_m.
  \end{equation}
\item The effective $n$th energy level is then obtained from
  the diagonal element $\Lambda_{nn}(t)$ via the well-known formula
  \begin{equation}\label{eq:effenergies}
    E_n^{\rm eff}(t) = \frac{1}{{\Delta}t}
    \ln\frac{\Lambda_{nn}(t)}{\Lambda_{nn}(t+{\Delta}t)}.
  \end{equation}
\end{enumerate}
We have used a timestep of ${\Delta}t=3a$ in our analysis. While the
choice of interpolators at the source is restricted to hexaquark
operators, we can probe a number of different operators at the sink
and study their relevance for determining the ground state.

Relying on non-hermitian correlator matrices makes it more difficult
to identify the ground state reliably, owing to the fact that the
diagonalization does not project exactly on the correlator
corresponding to the $n$th energy eigenstate. It is then not
guaranteed that the effective energies computed via
\eq{eq:effenergies} approach the asymptotic value monotonically from
above, since the statistical weights of different states are not
strictly positive.

In order to overcome this difficulty we have implemented distillation
and LapH smearing \cite{Peardon:2009gh}. Since this is a
timeslice-to-all rather than point-to-all method (see
Figure~\ref{fig:quark_lines_distillation}), it allows us to compute a
hermitian correlator matrix using the two-baryon operators listed in
Eqs.\,(\ref{eq:LL})--(\ref{eq:SS}), both in the center-of-mass frame
and for non-vanishing total momentum $\vec{P}$. The main objects that
we compute are the \emph{perambulator}, which is the projection of the
quark propagator onto the low modes of the spatial Laplacian, and the
\emph{mode triplets},
\begin{equation}
  T_{lnm}(t,\vec{p})=\sum_{\vec{x}} e^{-i\vec{p \cdot x}} \epsilon_{ijk}
  u^{(l,t)}_i(\vec{x}) u^{(n,t)}_j(\vec{x}) u^{(m,t)}_k(\vec{x}),
\end{equation}
where $u^{(l,t)}$ is the $l$th low mode on timeslice $t$. The
correlator matrix is then computed by contracting these two
objects. Further details of our implementation will appear in future
work; here we present initial results in the rest frame using
operators with both baryons at rest,
i.e.\ $\vec{p}_1=\vec{p}_2=\vec{0}$. For a hermitian
correlator matrix, the right and left eigenvectors are obviously
identical.

In the case of broken SU(3) symmetry, the diagonalization method also
allows us to associate a state with a particular multiplet, by
identifying which operators couple most strongly to it. Furthermore,
an essential feature of this method --- as will be shown in
Section~\ref{subsec:brokenSU3} --- is that it can identify the
presence of more than one state, even when the statistical signal is
too poor to distinguish the energies of those states.

We end this section with a remark on SU(3) flavor symmetry in our
setup, in which the strange quark is not present in the sea. In this
case flavor symmetry is realized by the graded group SU$(3|1)$. SU(3)
is a subgroup of SU$(3|1)$, and the flavor symmetry of our operator
construction is exact, even at the level of individual gauge
configurations. We have verified this by computing off-diagonal
correlators between different multiplets and found them to vanish
within errors.

\section{Energy levels}  \label{results}

\begin{figure*}
\includegraphics[width=0.75 \textwidth]{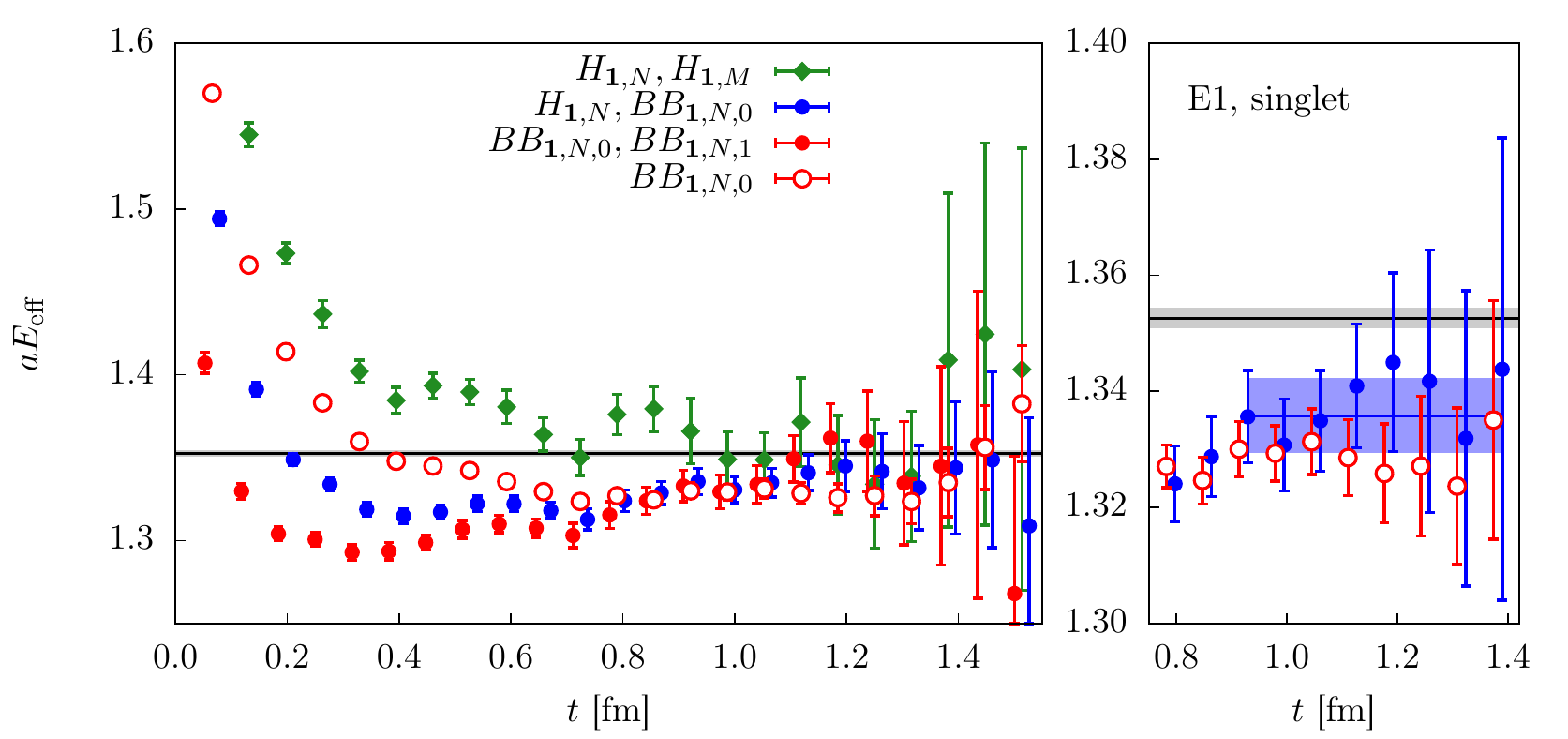}
\includegraphics[width=0.75 \textwidth]{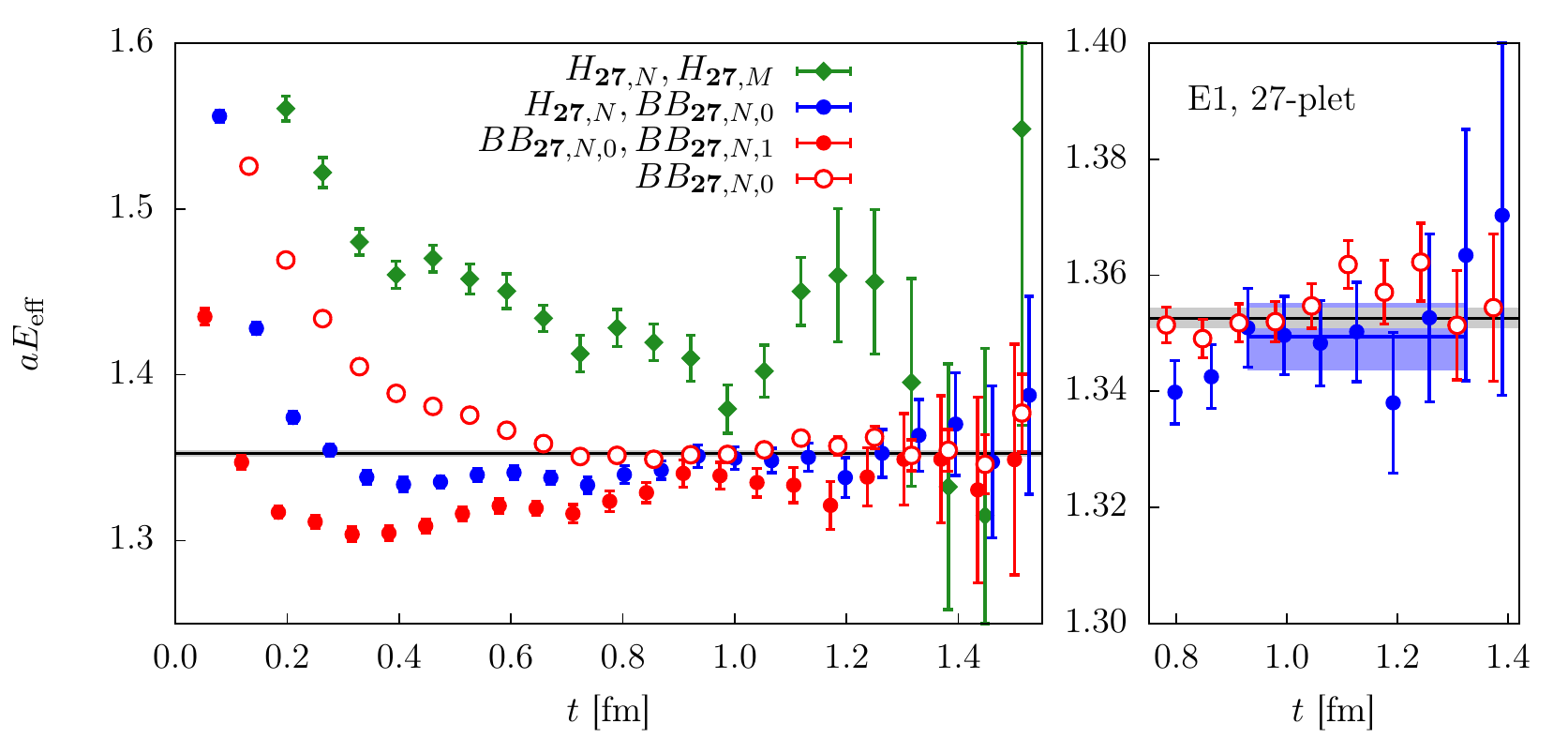}
\caption{\label{fig:E1-gevp} Ground-state effective energy for the
  singlet (top) and 27-plet (bottom) on ensemble E1. The legend
  indicates the operators used at the sink. Solid green diamonds
  denote the energies determined from the hermitian $2\times2$ GEVP,
  while the solid blue and red circles represent the results extracted
  from the non-hermitian $2\times2$ matrix. Open red circles
  correspond to the effective energy obtained from a single
  correlation function with a narrow-smeared hexaquark operator at the
  source and a two-baryon operator with $\vec{p}_1=\vec{p}_2=0$ at the
  sink. The plots on the right show the plateau region with the fitted
  energy levels. The horizontal line represents the value of
  $2m_\Lambda$, with the uncertainty denoted by the grey band.}
\end{figure*}

We now present our results for the ground-state energies in the
dibaryon channel, as determined using either point-to-all or
timeslice-to-all propagators on the ensembles listed in
Table~\ref{tab:lat_par1}. For completeness and further reference we
provide also the mass estimates for single baryons (see Tables
\ref{lambda} and \ref{octet}, as well as the corresponding effective
mass plots in Figures \ref{fig:octet} in Appendix~\ref{sec:appendix}.)

\subsection{Point-to-all propagators}\label{sec:analysis_point}

Correlators based on point-to-all propagators were computed for
ensembles A1, E1, E5 and N1. On all four ensembles we have computed
correlators in the rest frame and in three moving frames. Our main
findings are most easily explained for the data extracted from
ensembles E1 and E5 in the rest frame, which show the highest level of
statistical precision.

While A1, E1 and N1 realize an SU(3)-symmetric situation, SU(3)
symmetry is broken for ensemble E5. Note that the analysis for the
SU(3)-symmetric case is simplified, due to the fact that different
flavor multiplets (singlet, octet and 27-plet) cannot mix. Therefore,
we discuss the two cases separately.

\subsubsection{Analysis of the SU(3) symmetric case \label{subsec:SU3symm}}

As outlined in Section~\ref{sec:matrices} and illustrated in
Figure~\ref{fig:quark_lines_point}, our setup based on point-to-all
operators does not allow us to put two-baryon operators at the source.
Hence, we have constructed $2\times2$ correlator matrices using
hexaquark operators of the two different smearing types
($H_{\mathbf{1},N}$ and $H_{\mathbf{1},M}$ for the singlet case and
similarly for the 27-plet) at the source, while at the sink we have
used either hexaquark or two-baryon operators. In the latter situation
the resulting correlator matrix is non-hermitian, as described in
Section~\ref{sec:matrices}. After applying the appropriate
diagonalization procedure we have studied the behavior of the
effective energies defined according Eq.\,(\ref{eq:effenergies}), as a
function of the Euclidean time separation~$t$.

Results for the ground-state energy of the singlet channel on ensemble
E1 are shown in the top panel of Figure~\ref{fig:E1-gevp}. The
effective energy for the first excited state is too noisy to be
displayed in the plot. We find that the ``narrow'' smearing ($N$) is
the most effective in obtaining clean signals and therefore use only
narrow-smeared operators at the sink to determine our final results.

The green, blue and red filled symbols in Figure~\ref{fig:E1-gevp}
denote the energy levels extracted from different correlator matrices,
whose operator choice at the sink is described in the legend. The open
red circles denote the effective energy determined from a single
correlator, comprising a hexaquark operator at the source and a
two-baryon operator at the sink, with the two momenta each set to
zero. The horizontal line and grey band represent the energy level
corresponding to $2m_\Lambda$.

Our main observations are as follows: The effective energy determined
via the diagonalization of the hermitian $2\times 2$ hexaquark
correlator matrix approaches its plateau from above. In the plateau
region the data are noisy and compatible with the $\Lambda \Lambda$
threshold. By contrast, correlator matrices including at least one
two-baryon operator at the sink yield effective energies that are
below the threshold and which have smaller uncertainties. However,
care must be taken when deciding at which value of $t$ the asymptotic
behavior has been isolated. Owing to the non-hermitian setup, it is
possible that residual excited-state contributions enter the projected
ground-state correlator with negative weights, so that the plateau is
approached from below. This can also cause local minima to appear,
which could be difficult to distinguish from a true plateau. Indeed,
we see evidence for this behavior, with the energies showing a dip for
$t\approx0.4$\,fm before moving closer to the threshold for
$t\gtrsim0.9$\,fm. The effective energy determined from the mixed
single correlator approaches a plateau below the threshold, but
without showing any dip. We conclude that the ground-state effective
energy shows a consistent plateau for $t\gtrsim0.9$\,fm, which is
interpreted as the ground-state energy. The $2\times2$ hexaquark
correlator matrix yields consistent results for $t\gtrsim1.0$\,fm,
given the large statistical noise. 

Our estimate of the ground-state energy in the singlet channel on
ensemble E1 is obtained from a fit to the effective energy determined
from the diagonalization of the $2\times 2$ correlator matrix with one
hexaquark and one two-baryon operator at the sink. The panel on the
top right of Figure~\ref{fig:E1-gevp} shows a blow-up of the plateau
region, with the fitted value of the energy and error displayed as a
band across the fitting interval. Fitting the effective energy for
$t\gtrsim0.9$\,fm may still seem an optimistic choice, given that the
timeslices within the fit interval should satisfy $t>1/\Delta$, where
$\Delta$ is the energy gap to the first excited state. In a finite
volume with spatial length $L$, the energy gap is approximately given
by
\begin{equation}
\Delta\approx \frac{2}{E}\left(\frac{2\pi}{L}\right)^2\,,
\end{equation}
where $E$ is the energy of the ground state. Hence, for ensemble E1
one expects that $\Delta\approx170\,{\rm MeV}$, which translates into
$t\gtrsim 1.1$\,fm for ground-state dominance to be observed. While
this is inside the region where the signal is lost, we note that our
choice of fit interval is confirmed by our additional calculations
employing the distillation technique (see
Section~\ref{sec:analysis_distillation}), which yield statistically
much more precise results. In particular, we refer to
Figure~\ref{fig:point_vs_dist} below, which shows agreement between
the energy levels determined using point sources and distillation, for
Euclidean times $t\gtrsim0.9$\,fm. A thorough investigation of the
low-lying excitation spectrum, using the variational method in a fully
hermitian setup, will constitute a major part of our future work (see
Ref.~\cite{Hanlon:2018yfv} for preliminary results). By applying the
variational method, the gap to the nearest excited energy level is
increased, so that the timeslices inside the fit range easily satisfy
the condition $t> 1/\Delta$.

As a result of our fits, we find that the ground state in the singlet
channel lies below the energy of two non-interacting $\Lambda$
hyperons by~2.5 standard deviations. Numerical values for the fitted
energies are listed in Tables~\ref{lambda} and~\ref{singlet}, while
the results for the energy difference $2m_\Lambda-E$ are shown in
Table~\ref{singlet_diff}.

The qualitative features observed for the 27-plet are very similar to
the singlet channel (see bottom of Figure~\ref{fig:E1-gevp}): Whereas
the effective energy extracted from a correlator matrix constructed
from only hexaquark operators shows no sign of lying below the
$\Lambda\Lambda$ threshold, we find that using two-baryon operators
results in significantly lower values. Fitting the latter in the
region where $t>0.9$\,fm produces an estimate that lies within one
standard deviation below the threshold (see Tables~\ref{27plet}
and~\ref{27plet_diff}).

\subsubsection{Analysis of the SU(3)-broken case \label{subsec:brokenSU3}}

\begin{figure*}[ht]
\includegraphics[width=0.75\textwidth]{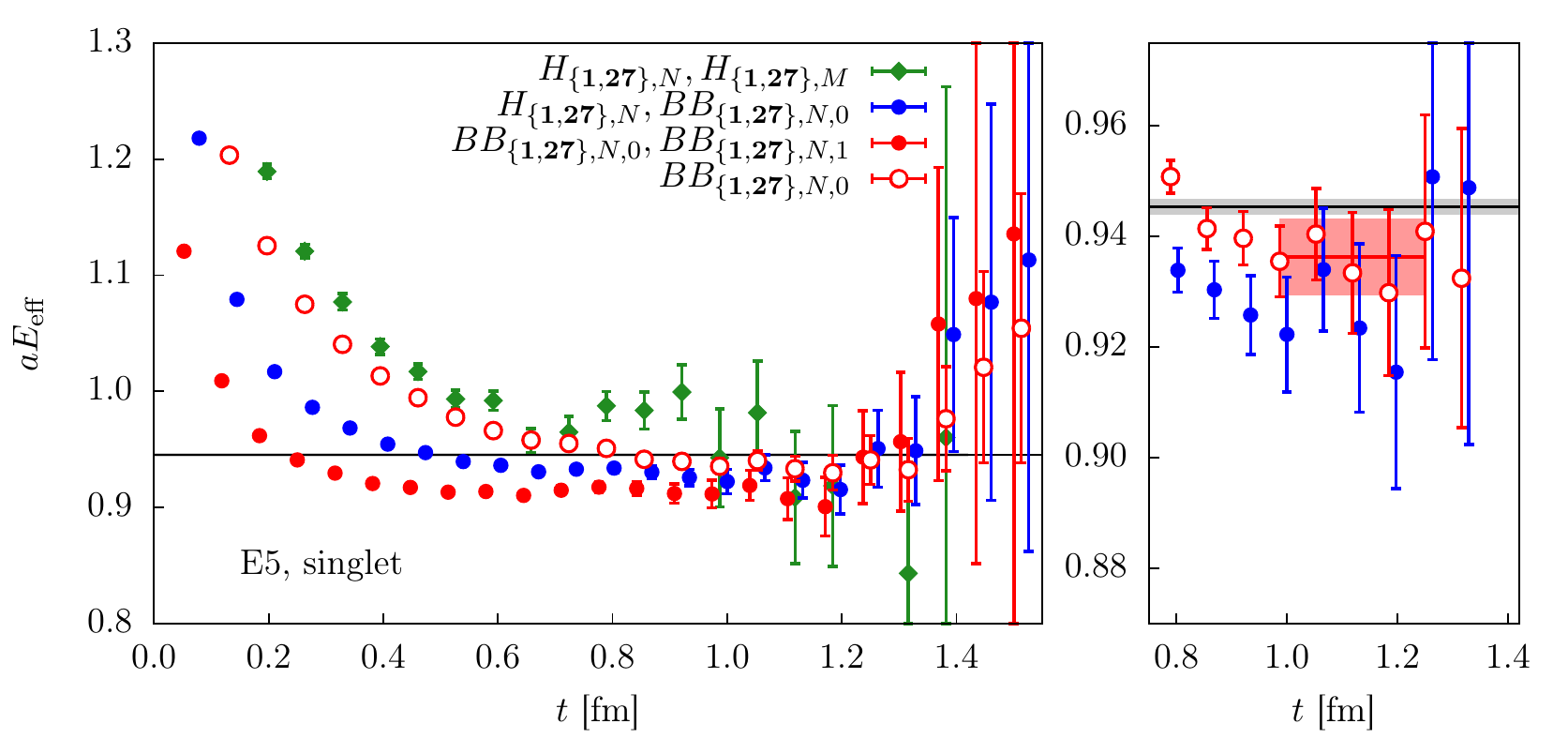}
\includegraphics[width=0.75\textwidth]{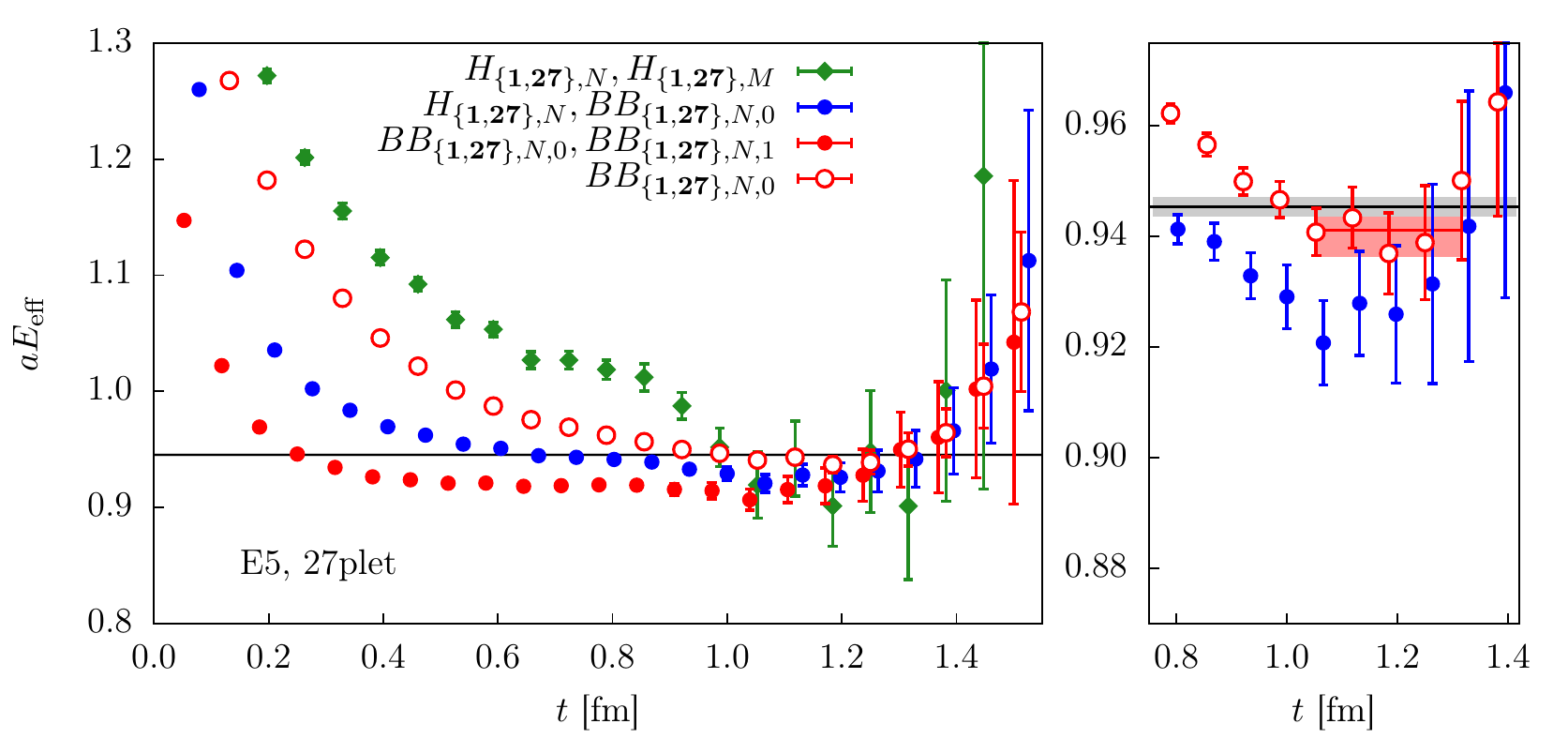}
\caption{\label{fig:E5-gevp} Ground-state effective energy for the
  singlet (top) and 27-plet (bottom) on ensemble E5 for which SU(3)
  symmetry is broken. The assignment of a particular multiplet to the
  energy levels was done on the basis of the dominant overlaps with
  the interpolating operator. The explanation of symbols is similar to
  that of Figure\,\ref{fig:E1-gevp}.}
\end{figure*} 

The breaking of SU(3) symmetry induces mixing among the relevant
multiplets (singlet, octet and 27-plet), which makes their
identification a more involved task. The appropriate strategy is to
construct correlator matrices from operators projected onto all
relevant flavor multiplets. We recall that the octet can only be
represented in terms of two-baryon operators, which, however, cannot
be placed at the source when using point-to-all
propagators. Therefore, the current analysis is focused only on the
singlet and 27-plet states. This deficiency will be rectified by the
calculation of hermitian correlator matrices using distillation,
described in the next subsection.

In the presence of mixing among different SU(3) multiplets, we have
constructed $4\times4$ correlator matrices from the same combinations
of operator types as in the SU(3)-symmetric case, while including both
projections onto the singlet and 27-plet. 

The matrix correlator and its diagonalization provides us with
important information for the interpretation of our results. First,
the amount of SU(3) breaking is small, as evidenced by the fact that
the geometric mean of the flavor off-diagonal correlators is less than
0.5\% of that of the corresponding flavor-diagonal ones. Second, the
diagonalization provides clear evidence for the existence of two
different low-lying states, although it is not possible to resolve
their energy difference with the current statistics. Furthermore,
states corresponding to different flavor multiplets can be identified
with the help of the eigenvectors computed via the diagonalization
procedure. We find that the eigenvectors corresponding to the two
low-lying states are dominated by operators in a single multiplet. In
this way we can unambiguously assign the energy levels to a particular
flavor multiplet.

\begin{figure*}[ht]
\leavevmode
\includegraphics[width=0.45\textwidth]{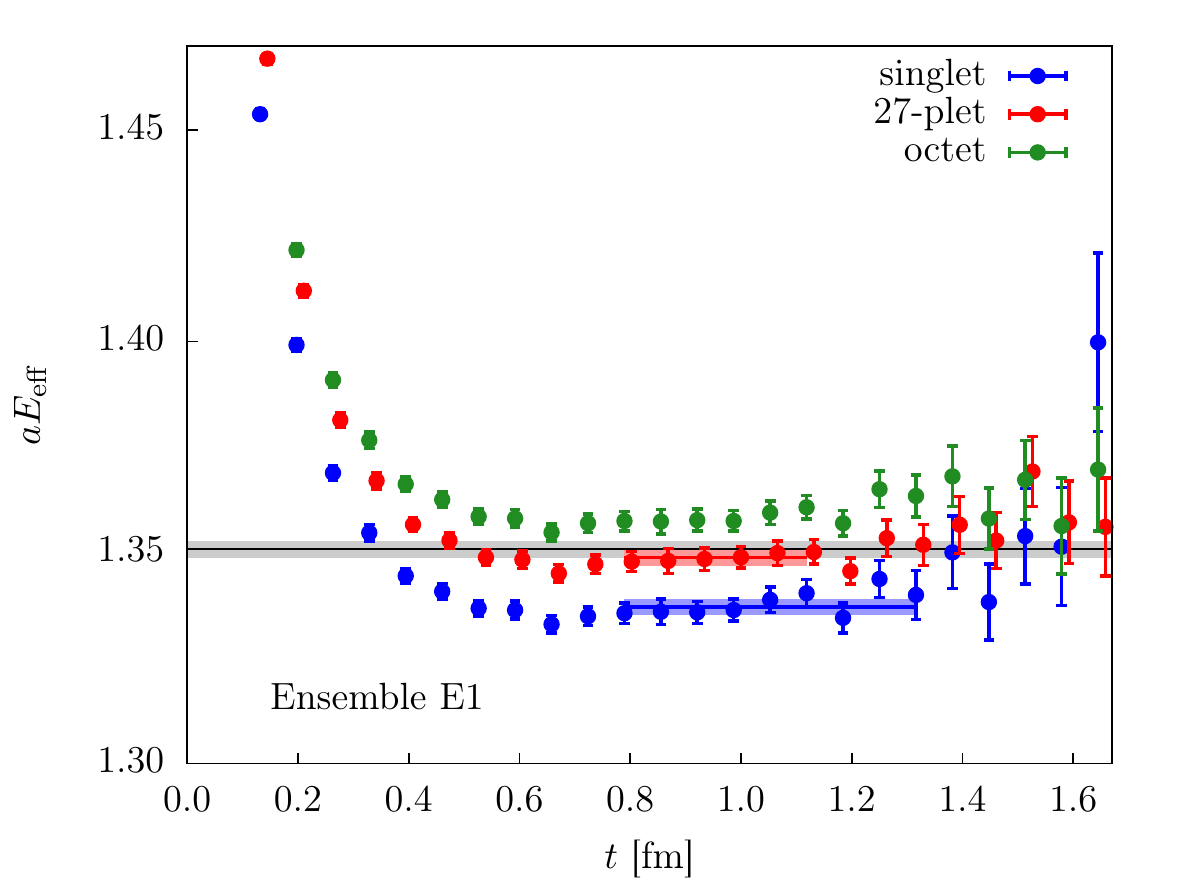}
\hfill
\includegraphics[width=0.45\textwidth]{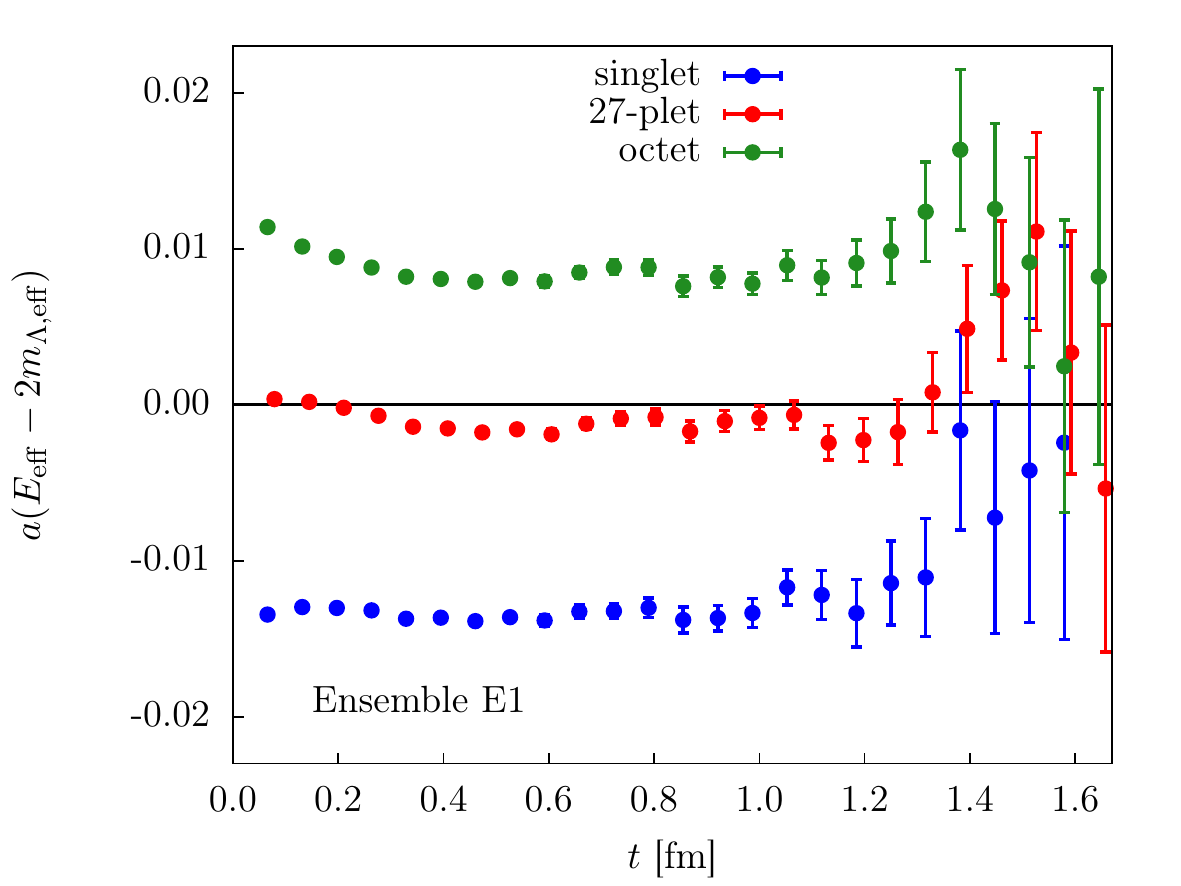}
\includegraphics[width=0.45\textwidth]{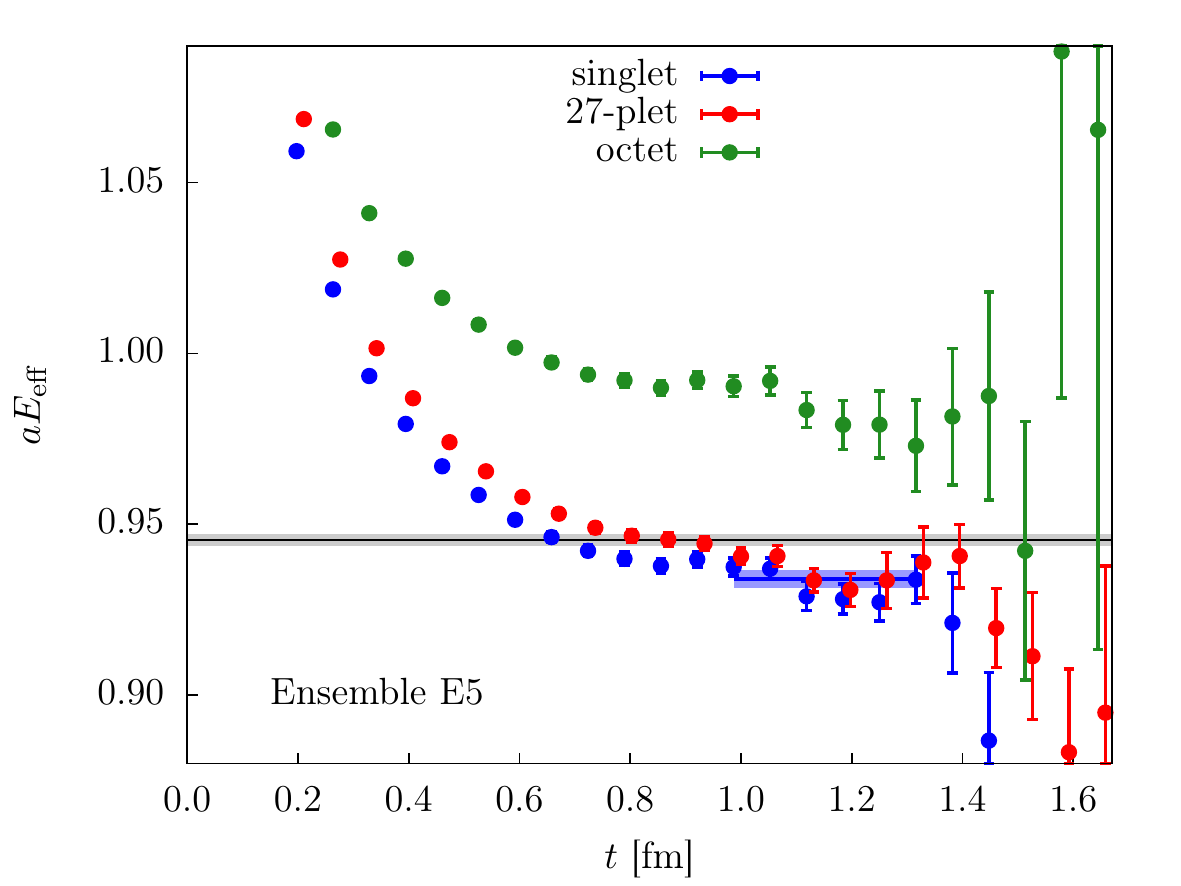}
\hfill
\includegraphics[width=0.45\textwidth]{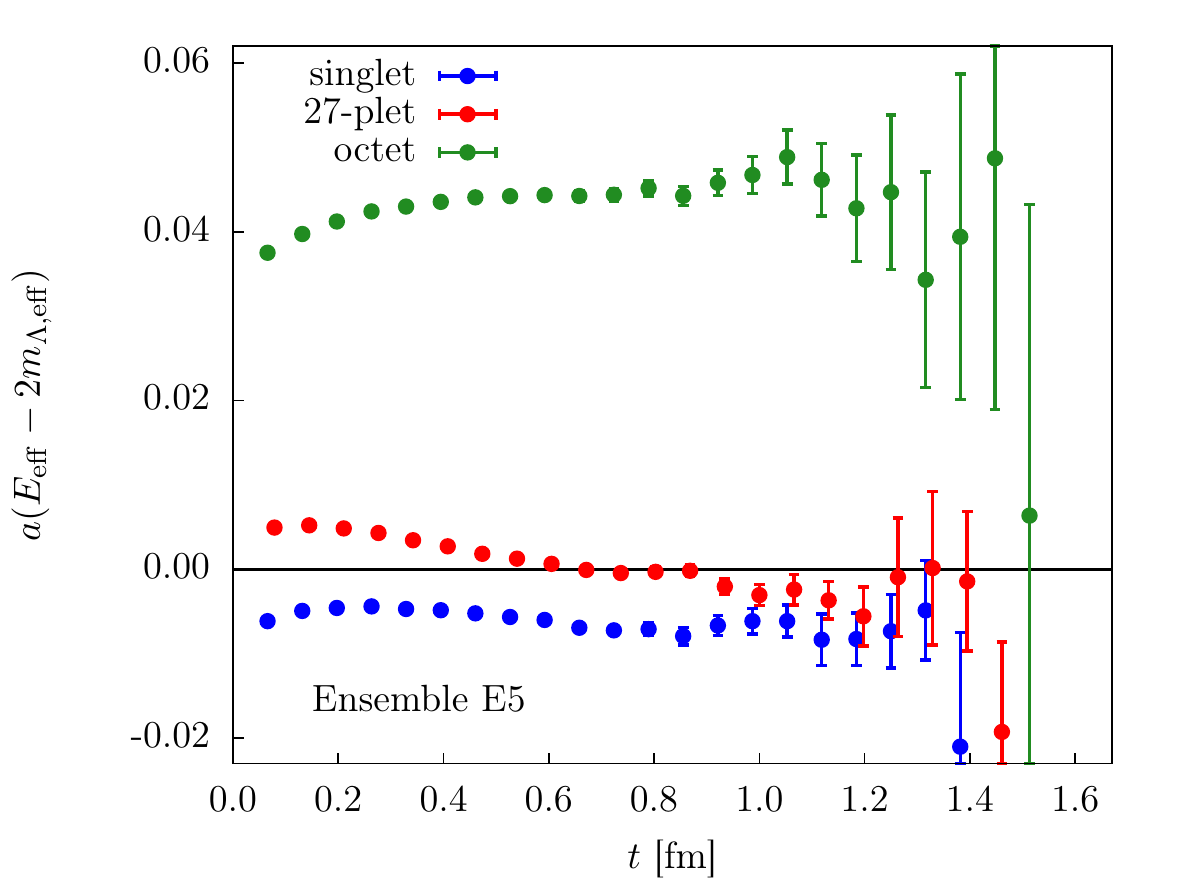}
\caption{\label{fig:distill} Top: Effective energies for different
  multiplets for the SU(3)-symmetric case (ensemble E1). The right
  panel shows the effective energy difference relative to the
  $\Lambda\Lambda$ threshold. Bottom: energy levels and relative
  difference for the SU(3)-broken case (ensemble E5). As in
  Figure~\ref{fig:E5-gevp}, the assignment of flavor multiplets to the
  energy levels was done by means of the dominant overlaps with the
  interpolating operator. All results have been obtained using
  distillation. Colored bands indicate the fitted values across the
  relevant time interval.}
\end{figure*}

Results for the effective energies for the SU(3) singlet and 27-plet
are shown in the top and bottom panels of Figure~\ref{fig:E5-gevp},
respectively. Data points shown in a given color correspond to a
particular choice of sink operators, as indicated in the legend. For
both the singlet and the 27-plet, results are similar to what was
found for ensemble E1. Effective energies from diagonalizing
$4\times4$ correlator matrices that include two-baryon operators at
the sink (filled red and blue circles) dip below the $\Lambda\Lambda$
threshold before $t=0.5$\,fm; however, they are in disagreement with
each other for $t\lesssim1.0$\,fm. Statistically, the most precise
signal is obtained by considering a $2\times2$ correlator matrix,
composed of ``narrow'' smeared hexaquark interpolating operators for
the singlet and 27-plet at the source and a corresponding set of
two-baryon operators with both momenta set to zero at the sink. The
corresponding effective energy is represented by the open red circles
in Figure~\ref{fig:E5-gevp}. After fitting the effective energy to a
constant in the interval $t=1.0 - 1.3$\,fm we obtain the estimates
represented by the red bands in the right panels. We find that the
fitted ground-state energies of the singlet and 27-plet are of the
order of one standard deviation below the $\Lambda\Lambda$ threshold
(see Tables~\ref{lambda}\,--\,\ref{27plet_diff}). Note that this
analysis is in slight tension with the effective energy from a
$4\times 4$ correlator matrix with only two-baryon operators at the
sink (filled red circles), meaning that the uncertainty may be
somewhat underestimated.

As discussed in Section~\ref{sec:matrices} we have computed
correlation matrices in frames with total momentum
$\vec{P}=2\pi\vec{D}/L$ with $\vec{D}^2=0, 1, 2, 3$. Estimates for the
corresponding ground-state energies are listed in
Tables\,\ref{singlet}\,--\,\ref{27plet_diff}. The combined results in
all frames have been used to obtain information on the scattering
phase shifts and binding energies in infinite volume. A detailed
discussion is deferred to Section~\ref{sec:finite_volume}.

\subsection{Dibaryon analysis with distillation}\label{sec:analysis_distillation}

The distillation technique enables the use of two-baryon operators at
both the source and the sink. This results in a hermitian correlator
matrix and allows us to properly account for the mixing between the
singlet, octet and 27-plet states under SU(3) breaking. Within this
setup we do not use hexaquark operators. For our initial study, we
have computed correlation functions on ensembles E1 and E5 in the rest
frame only, which allows for a direct comparison with results from the
previous subsection.

As described in Section~\ref{sec:interpolators}, we start from the
two-baryon interpolating operators of
Eqs.~(\ref{eq:LL})--(\ref{eq:SS}), projected on
$\vec{p}_1=\vec{p}_2=0$. Using the transformation of Appendix~B in
Ref.\,\cite{Inoue:2010hs} we then perform the rotation to the singlet,
octet and 27-plet, which yields the operators $BB_{\mathbf{1},L,0}$,
$BB_{\mathbf{8},L,0}$ and $BB_{\mathbf{27},L,0}$, respectively. In
each channel we compute the corresponding correlation functions and
determine the effective energies. Results for the effective energy on
E1 are shown in the top left panel of Figure~\ref{fig:distill}. In all
three channels, the effective energy approaches a plateau
monotonically from above and do not show any of the irregularities
observed in the non-hermitian setup based on point-to-all
propagators. The top right panel of Figure~\ref{fig:distill} shows the
effective energy difference $E_{\text{eff}}-2m_{\Lambda,\text{eff}}$.

The plot demonstrates that the statistical precision in each of the
three individual ($1\times1$) correlators is sufficient to separate
the energy levels corresponding to the singlet, 27-plet and octet
channels (in ascending order). The fact that the effective energy
differences are nearly constant over time indicates that excited-state
contributions in the dibaryon correlation functions are very similar
to those in the single $\Lambda$ correlator. Still, it is important to
determine the energy difference in the regime where the individual
correlators have reached their respective asymptotic behavior. In the
singlet channel (color-coded in blue), the effective energy in the
plateau region lies significantly below the non-interacting $\Lambda
\Lambda$ threshold. The energy level of the 27-plet is closer to the
threshold, while the octet state lies significantly above.

\begin{figure}[t]
\includegraphics[width=0.98\columnwidth]{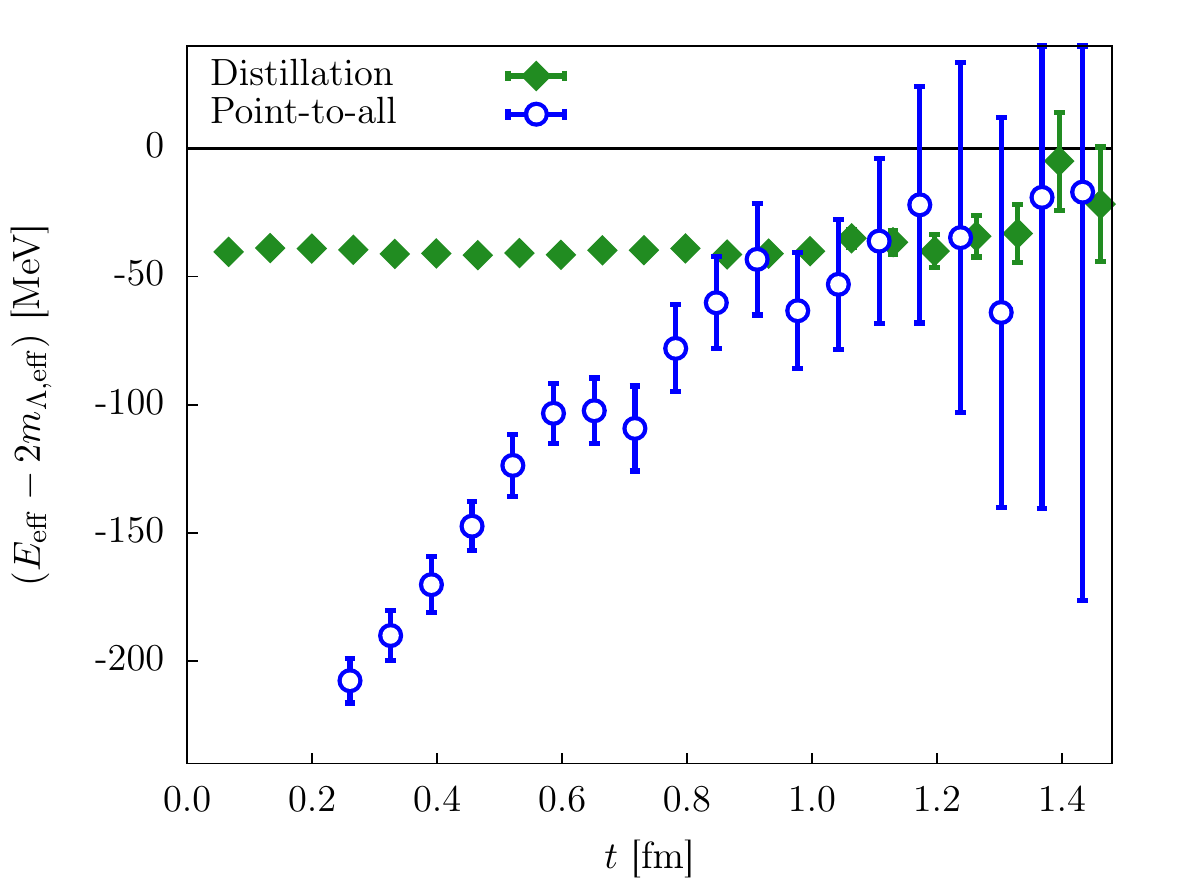}
\caption{\label{fig:point_vs_dist} Comparison of the effective energy
  difference between the singlet ground state and two non-interacting
  $\Lambda$s, computed with point-to-all propagators and distillation,
  respectively, for ensemble~E1. Open blue circles derive from the
  same non-hermitian $2\times2$ correlator matrix that was used to
  determine the ground state energy in the singlet channel,
  represented by the blue points in the top panel of
  Figure~\ref{fig:E1-gevp}.}
\end{figure}

The results on E5 are shown in the lower half of
Figure~\ref{fig:distill}, with color coding identical to that of
ensemble~E1. Due to the breaking of SU(3) symmetry, the different
multiplets cannot be treated independently anymore and it is here that
the inclusion of the octet operator at the source and sink allows for
a consistent treatment of the relevant flavor multiplets. In order to
take account of the mixing of different multiplets, we compute a
$3\times3$ correlator matrix using the operators
$BB_{\mathbf{1},L,0}$, $BB_{\mathbf{8},L,0}$ and
$BB_{\mathbf{27},L,0}$, which is then subjected to the diagonalization
procedure. The resulting energy levels are identified with a
particular multiplet according to the overlaps with the interpolating
operator. The effective energy of the singlet (blue circles) is below
the $\Lambda\Lambda$ threshold and, in contrast to the case of
point-to-all propagators, can be clearly distinguished from the energy
corresponding to two non-interacting $\Lambda$s. The distillation
technique allows us to distinguish the singlet and 27-plet as two
distinct states with our statistics. The octet, by contrast, lies far
above the threshold. For our final estimates of the energies of the
different states in the SU(3)-symmetric situation, we have fitted the
effective energy to a constant for $t$ starting from 0.8\,fm. In the
SU(3)-broken situation, the plateau starts later and we have fitted
from 1.0\,fm.

It is instructive to compare the energy levels computed using
point-to-all propagators to the results extracted using the
distillation technique. In Figure~\ref{fig:point_vs_dist} we show the
effective energy gap for the singlet ground state computed on
ensemble~E1. The plot demonstrates the dramatic improvement in the
overall precision provided by the distillation technique in
conjunction with LapH smearing: The effective energy gap computed
using distillation can be clearly distinguished from zero owing to its
tiny statistical errors. Furthermore, it exhibits a flat behavior
across a wide range of Euclidean times. While the results obtained
using point-to-all propagators are consistent with distillation for $t
\gtrsim 0.9$\,fm, the larger statistical noise makes it more difficult
to determine a non-vanishing energy gap. Also, one does not profit
from the partial cancellation of statistical noise in the difference
$E_{\rm eff}-2m_{\Lambda,{\rm eff}}$ when using point sources compared
to using distillation. As with the point-source method, some caution
is required due to the small energy gap to the first excited state. If
we assume that the states resemble the noninteracting ones, then the
two-baryon operator with both baryons at rest will couple strongly to
the ground state and poorly to the elastic excited states; this could
explain the long plateau in Figure~\ref{fig:point_vs_dist}. In fact,
the length of this plateau is comparable to the inverse energy gap
$\Delta^{-1}\approx 1.1$\,fm.

We end this section with a discussion of the relative cost of
calculations based on point-to-all propagators and distillation. A
simple and fairly accurate cost estimate is provided by the number of
propagator solves per configuration. For point sources we have to
perform
\begin{equation}
   N_{\text{pt}}= N_{\text{flav}}\cdot  N_{\text{src}} \cdot
     N_{\text{smear}} \cdot 12 
\end{equation}
inversions, where $N_{\text{flav}}$ denotes the number of different
quark flavors, $N_{\text{src}}$ is the number of sources, and the
number of different smearings is given by $N_{\text{smear}}$. For
distillation the number of propagator solves is given by
\begin{equation}
   N_{\text{dist}}= N_{\text{flav}}\cdot  N_{\text{tsrc}} \cdot
     N_{\text{LapH}} \cdot 4,
\end{equation}
where $N_{\text{tsrc}}$ denotes the number of timeslices used to
compute timeslice-to-all propagators, and $N_{\text{LapH}}$ is the
number of low-lying modes of the spatial Laplacian. Using the
information from Table~\ref{tab:lat_par1} and the fact that we have
used $N_{\text{LapH}}=56$ modes, we find $N_{\text{pt}}=3072$ and
$N_{\text{dist}}=1792$ for ensemble E1. For E5 the above expressions
evaluate to $N_{\text{pt}}=1536$ and $N_{\text{dist}}=1792$. We
conclude that the better data quality of the distillation technique is
achieved for comparable or even lower cost. For a more precise cost
comparison one should also take into account that the cost of
contractions in the distillation approach is typically larger. The
associated computational overhead may become significant on large
volumes. We will present a more thorough discussion of this issue in a
future publication.

\section{\label{sec:finite_volume}Finite volume analysis}

The finite-volume rest frame energy level $E$, when below the
two-particle threshold, provides a na\"{\i}ve estimate of the mass of the
bound state. However, this can suffer from significant finite-volume
effects that are asymptotically only suppressed as $e^{-\kappa L}$,
where $\kappa$ is the binding momentum defined via
$E = 2\sqrt{m_\Lambda^2 - \kappa^2}$. In addition, there can be an
energy level below threshold without a bound state, in which case it
has a power-law dependence on $L$ that depends on the scattering
length.

For a more careful study of the presence of a bound state and its
mass, we turn to L\"uscher's finite volume quantization
condition~\cite{Luscher:1990ux} and its extension to moving
frames~\cite{Rummukainen:1995vs}. Below the three-particle threshold,
this is a relation between the two-particle scattering amplitude and
the finite-volume energy levels. In the case of one pair of identical
particles scattering in the $S$ wave, if we ignore the influence of
higher partial waves due to the breaking of rotational symmetry, the
condition takes the form
\begin{equation}
  p\cot\delta(p) = \frac{2}{\sqrt{\pi}L\gamma} Z_{00}^{\mathbf{D}}\left(1,\left(\frac{pL}{2\pi}\right)^2\right),
\end{equation}
where $p$ is the scattering momentum satisfying
$E_\text{cm}\equiv\sqrt{E^2-\vec{P}^2}=2\sqrt{m_\Lambda^2+p^2}$,
$\delta(p)$ is the scattering phase shift, $\gamma=E/E_\text{cm}$ is
the moving-frame boost factor, and
$Z_{00}^{\mathbf{D}}$ is a generalized zeta function defined in
Ref.~\cite{Rummukainen:1995vs}. Our numerical implementation of the
zeta function is based on Ref.~\cite{Gockeler:2012yj}.

As recently reviewed in Ref.~\cite{Iritani:2017rlk}, a bound state,
corresponding to a pole in the scattering amplitude on the real $p^2$
axis below zero, is determined by the condition
$p\cot\delta(p)=-\sqrt{-p^2}$. That reference also provides a check
that can be applied to lattice data: at the pole, the slope of
$p\cot\delta(p)$ (versus $p^2$) must be smaller than that of
$-\sqrt{-p^2}$.

For small $p^2$, the phase shift can be described by the effective
range expansion. We use the first two terms,
\begin{equation}
  p\cot\delta(p) = -\frac{1}{a_0} + \frac{r_0}{2}p^2,
\end{equation}
where $a_0$ is the scattering length and $r_0$ is the effective
range. Fitting this to lattice data is not a simple linear fit, since
$p\cot\delta(p)$ and $p^2$ are not independent variables, being
related by the quantization condition. We choose to fit to the squared
scattering momentum $p^2$ determined from each lattice energy, in a
manner similar to what was done for fitting to the lattice energies in
Ref.~\cite{Erben:2017hvr}. Fitting to the momentum rather than energy
benefits from cancellations of statistical uncertainties that are
correlated between $E$ and $m_\Lambda$.  Given a set of fit parameters
$(a_0,r_0)$, in each frame the fit momentum is determined by finding
the solution to the quantization condition that is nearest to the
corresponding momentum determined from the lattice calculation,
i.e.\ by numerically solving
\begin{equation}
\frac{2}{\sqrt{\pi}L\gamma} 
Z_{00}^{\mathbf{D}}\left(1,\left(\frac{pL}{2\pi}\right)^2\right)
 = -\frac{1}{a_0} + \frac{r_0}{2}p^2.
\end{equation}
We only consider the ensembles with SU(3) symmetry, since otherwise
this is a much more complicated coupled-channel ($\Lambda\Lambda$,
$N\Xi$, and $\Sigma\Sigma$) system.

\begin{figure}
  \centering
  \includegraphics[width=\columnwidth]{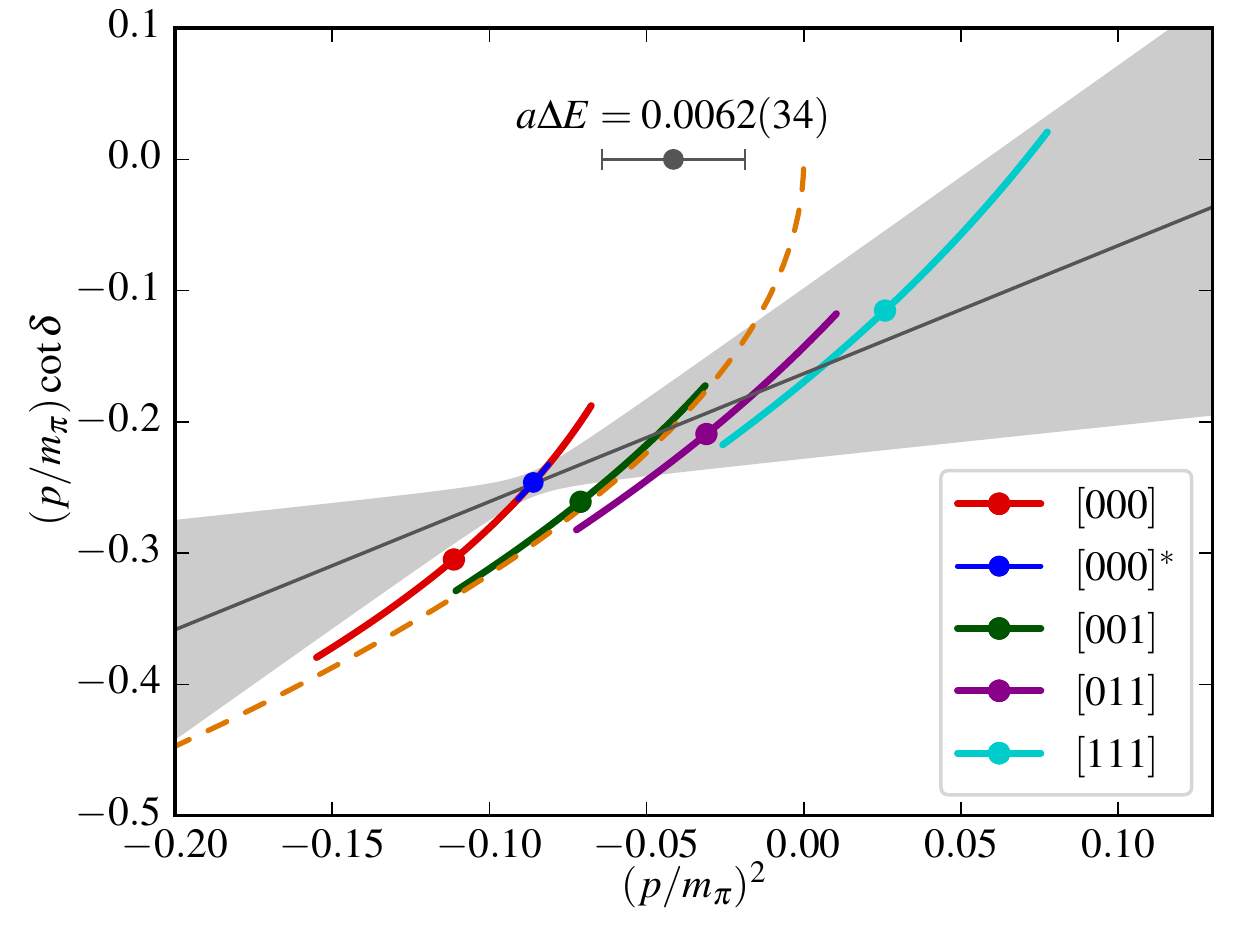}
  \caption{SU(3) singlet scattering phase shift on ensemble E1. The
    legend indicates the moving frame $\vec{D}=\tfrac{L}{2\pi}\vec{P}$;
    the data point labeled $[000]^*$ was obtained using distillation and the
    others were obtained using point-source data. The grey line
    and its error band indicate the effective-range-expansion fit, and
    the orange dashed curve corresponds to
    $p\cot\delta=-\sqrt{-p^2}$. The horizontal error bar shows the
    intersection between the grey line and the orange dashed curve,
    which is translated to a binding energy in the label above it.}
  \label{fig:E1_singlet_phase_shift}
\end{figure}

We are only able to obtain a reliable fit in the flavor singlet sector
on ensemble E1, where the precise energy level in the rest frame from
the distillation method provides a stronger constraint than the other
data. This is shown in Figure~\ref{fig:E1_singlet_phase_shift}; we
find the scattering length to be 1.3(5)~fm and the effective range
0.4(3)~fm. There is a clear intersection with the bound-state curve,
which has a slope with the correct behavior, indicating the presence
of a bound $H$ dibaryon. This intersection yields a binding energy of
$\Delta E=19(10)$~MeV, somewhat smaller in magnitude than the
na\"{\i}ve value obtained from the rest-frame energy difference
$2m_\Lambda-E$.

\begin{figure}
  \centering
  \includegraphics[width=\columnwidth]{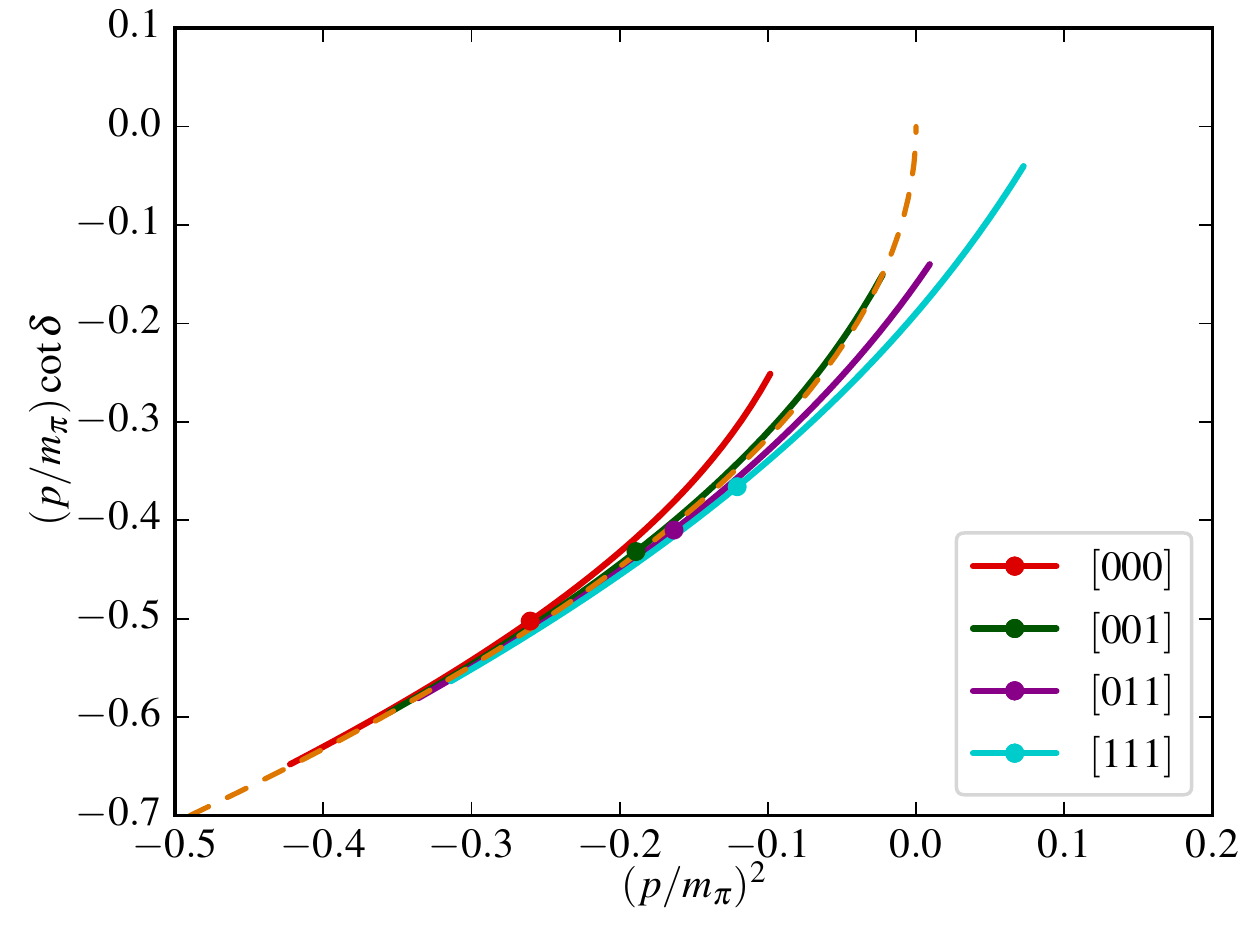}
  \caption{SU(3) singlet scattering phase shift on ensemble A1. The
    legend indicates the moving frame
    $\vec{D}=\tfrac{L}{2\pi}\vec{P}$ and the
    orange dashed curve corresponds to $p\cot\delta=-\sqrt{-p^2}$.}
  \label{fig:A1_singlet_phase_shift}
\end{figure}

For the ensembles A1 and N1, we are unable to obtain reliable
fits. However, in the flavor singlet sector, the data in the rest
frame and in the higher moving frames sit on opposite sides of the
bound-state curve, which indicates that there will be an intersection
between it and the phase shift; see
Figure~\ref{fig:A1_singlet_phase_shift}. For the frame $\vec{D}^2=1$,
the point --- together with its error along the quantization curve ---
is almost on top of the bound-state curve, and therefore we make a
conservative estimate of the binding energy. To this end we consider
the interval defined by the maximum and minimum values of
$E_\text{cm}$ deduced from the frames with $\vec{D}^2=0, 1$. The
midpoint of this interval is identified with the central value of the
binding energy, while the $1\sigma$ error is defined as the difference
with the upper and lower bounds. For N1, we get $a\Delta E=0.016(13)$,
or 65(53)~MeV. For A1, we get $a\Delta E=0.028(25)$, or 73(66)~MeV.

\begin{figure}
  \centering
  \includegraphics[width=\columnwidth]{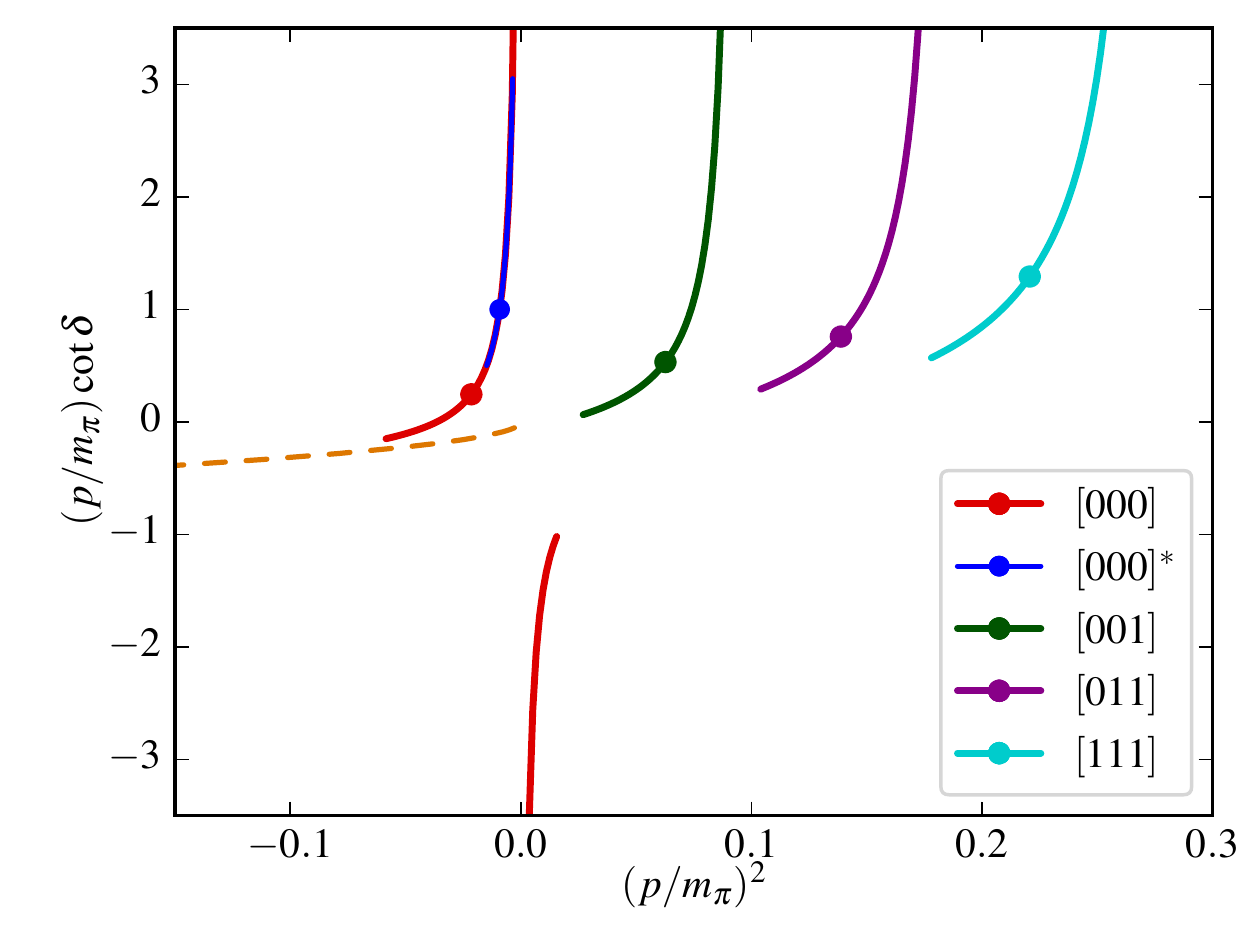}
  \caption{SU(3) 27-plet scattering phase shift on ensemble E1. The
    legend indicates the moving frame
    $\vec{D}=\tfrac{L}{2\pi}\vec{P}$; the data point labeled $[000]^*$
    was obtained using distillation and the others were obtained using
    point-source data. The orange dashed curve corresponds to
    $p\cot\delta=-\sqrt{-p^2}$.}
  \label{fig:E1_27plet_phase_shift}
\end{figure}

In the 27-plet sector, our data are generally not precise enough to
distinguish the energy levels from the noninteracting ones; this means
that the phase shift is consistent with zero. On E1, where
distillation provides a relatively precise value, the rest-frame
energy is slightly below threshold. Using the quantization condition,
we find that just below threshold, $p\cot\delta$ is probably positive;
see Figure~\ref{fig:E1_27plet_phase_shift}. This means that a bound
state (which would be a dineutron) is unlikely, however an
intersection with $p\cot\delta=+\sqrt{-p^2}$, which corresponds to a
virtual bound state (i.e., a pole on the unphysical sheet), is
possible. If we neglect the dependence on $p^2$, the distillation
energy level implies a scattering length
$a_0=-0.2_{-0.2}^{+0.1}$~fm. However, note that the quantization
condition is very nonlinear and at 3$\sigma$ the statistical
uncertainty on $a_0$ becomes as large as ${}^{+0.3}_{-1.2}$, assuming
a symmetric Gaussian uncertainty on the $p^2$ determined from the
energy level.

\section{Discussion and Conclusion} \label{sec:conclusions}

We have presented a detailed study of the spectrum of the $H$~dibaryon
in the SU(3) flavor-symmetric and broken cases, using lattice QCD
simulations with dynamical up and down quarks and a quenched strange
quark. We have analyzed the efficiency of different interpolating
operators in the determination of hadronic finite volume energy
eigenvalues via variational analysis.

Our findings indicate that point-like hexaquark operators provide a
poor overlap onto the SU(3) singlet ground state. This becomes evident
through the slow convergence to the ground state plateau seen in the
time dependence of the lowest energy eigenvalue. The inclusion of
bilocal two-baryon operators at the sink improves the overlap
considerably, as indicated by an earlier onset of the plateau and a
stronger statistical signal. However, in a setup based on point-to-all
propagators, the improvement comes at the expense of having to deal
with non-hermitian correlator matrices. As a consequence, effective
energies determined from the eigenvalues are not guaranteed to
approach their asymptotic behavior from above. In this way an
additional systematic is introduced, as the onset of the plateau is
obscured or can be misidentified.

\begin{figure}[t]
\includegraphics[width=0.98\columnwidth]{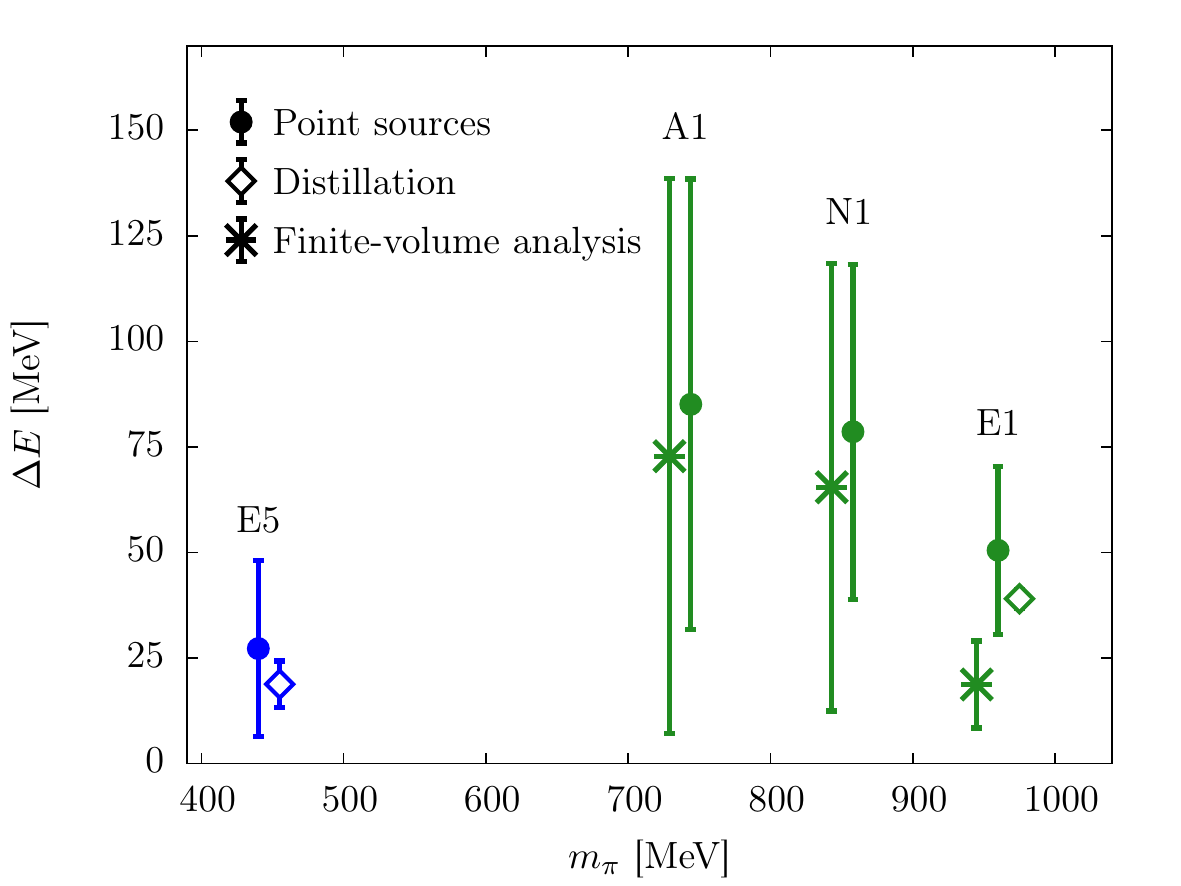}
\caption{\label{fig:summary} Summary of our results for the binding
  energy of the $H$~dibaryon. Green and blue colors refer to the
  SU(3)-symmetric and broken cases, respectively. Full circles and
  open diamonds represent results obtained in the rest frame using
  point sources and distillation, respectively. Crosses denote
  estimates for the binding energy extracted from the finite-volume
  analysis of results in different frames as described in
  Section~\ref{sec:finite_volume}.}
\end{figure}

A fully hermitian setup with bilocal two-baryon operators at both the
source and sink can be realized with the help of the distillation
technique which allows for the calculation of timeslice-to-all
propagators. In our study of the SU(3) symmetric and broken cases we
have been able to determine the energy levels reliably and with good
statistical precision for the flavor singlet, 27-plet and octet.

In particular, we could identify a clear gap between the ground-state
singlet energy level and the $2m_\Lambda$ threshold, which suggests
that there is a bound $H$~dibaryon. The energy difference provides a
na{\"\i}ve estimate of the binding energy, ${\Delta}E\equiv
2m_\Lambda-E$. For ensembles E1 and E5, which realize the SU(3)
symmetric and broken situations respectively, the estimates are
\begin{align}
\text{E1:}\quad & \Delta E = 39.0\pm2.2\,\text{MeV},\quad
m_\pi=960\,\text{MeV}, \label{eq:bind_E1dist} \\
\text{E5:}\quad & \Delta E = 18.8\pm5.5\,\text{MeV},\quad
m_\pi=440\,\text{MeV}. \label{eq:bind_E5dist} 
\end{align}
However, recalling that finite-volume effects are asymptotically
suppressed as $e^{-{\kappa}L}$, where $\kappa$ is the binding
momentum, the na{\"\i}ve estimates of the binding energy in
Eqs.~(\ref{eq:bind_E1dist}) and~(\ref{eq:bind_E5dist}) may not be very
reliable, given that ${\kappa}L$ evaluates to 2.99(8) for E1 and
1.74(25) for E5.

In addition, we have applied L\"uscher's finite-volume quantization
condition to determine the scattering phase shift, which we use to
obtain a more reliable estimate of the binding energy. Including data
from the rest frame as well as several moving frames in the
finite-volume analysis of the singlet case results in a shallower
binding energy compared to the na\"{\i}ve estimate:
\begin{equation}
\label{eq:FVgap}
\text{E1:}\quad \Delta E = 19\pm10\,\text{MeV},\quad m_\pi=960\,\text{MeV}.
\end{equation}
Repeating the analysis for the SU(3) 27-plet is made more difficult
through the relative closeness of the energy levels to the
non-interacting levels, and the same is true in the case of broken
SU(3)-flavor symmetry. We expect the situation to improve in our
ongoing work, since we will obtain more precise results by also using
distillation in the moving frames.

\begin{figure}[t]
\includegraphics[width=0.98\columnwidth]{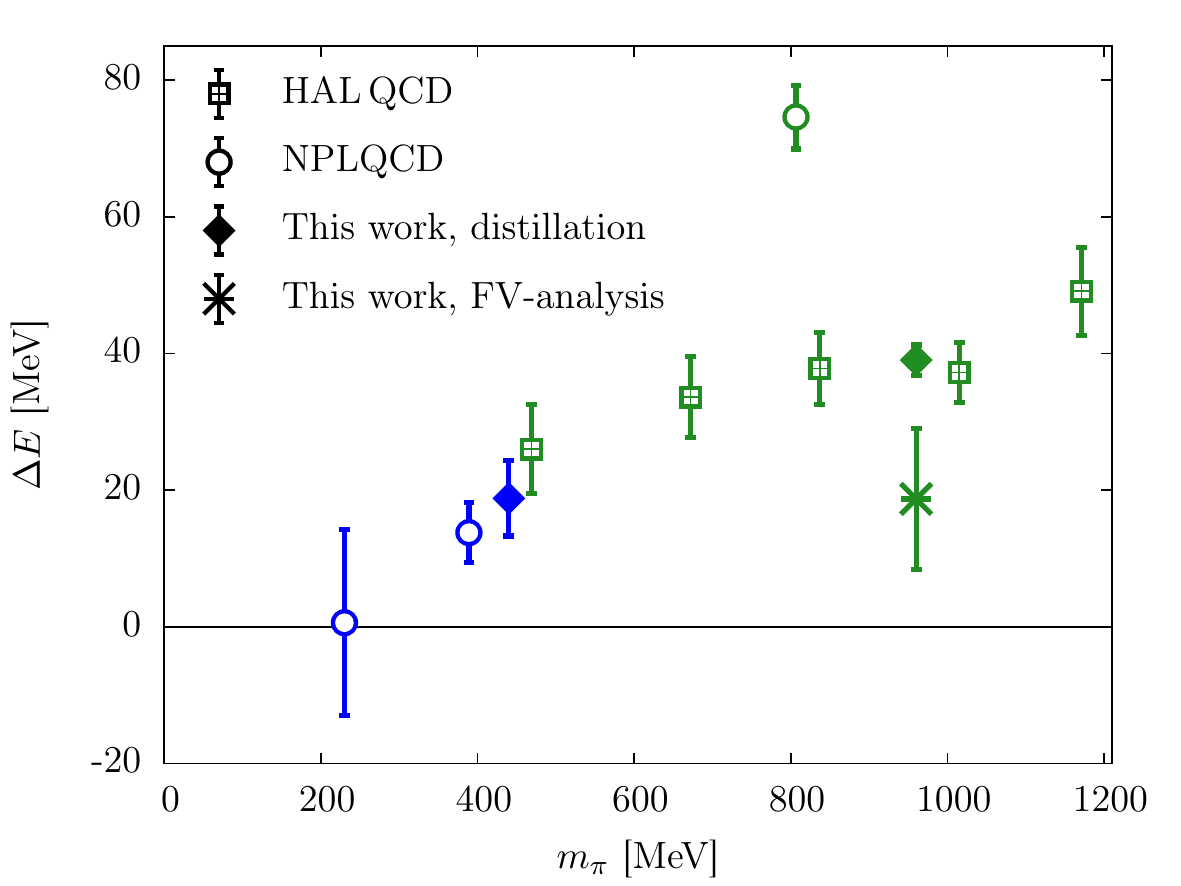}
\caption{\label{fig:comparison} Comparison of our results in
  Eqs.~(\ref{eq:bind_E1dist})--(\ref{eq:FVgap}) to the estimates
  quoted by NPLQCD \cite{Beane:2010hg,Beane:2011zpa,Beane:2012vq} and
  HAL~QCD \cite{Inoue:2010es,Inoue:2011ai}. Green and blue symbols
  refer to the SU(3)-symmetric and broken cases, respectively. The
  data point marked by a star denotes the result in infinite volume.}
\end{figure}

In Figure~\ref{fig:summary} we show a compilation of our results for
the binding energy on all our ensembles, plotted against the pion
mass. The comparison with the estimates of the
NPLQCD\,\cite{Beane:2010hg,Beane:2011zpa,Beane:2012vq} and
HAL~QCD\,\cite{Inoue:2010es,Inoue:2011ai} collaborations is made in
Figure~\ref{fig:comparison}. In the SU(3)-symmetric case we find that
our estimate is considerably smaller than the result quoted by NPLQCD
at a similar value of the pion mass \cite{Beane:2010hg}. Potentially,
uncontrolled systematics such as the incorrect identification of the
plateau, quenching of the strange quark or finite-volume effects could
be the source of this discrepancy. We will address these issues in a
future publication based on ensembles with $N_{\text{f}}=2+1$ flavors
of dynamical quarks. 

Our findings suggest that the combination of distillation and
L\"uscher's finite-volume formalism will allow for a considerably
improved calculation of the binding energy. Thus, there are good
prospects for a reliable determination of this quantity at the
physical point.


\begin{acknowledgments}
We are grateful to Maxwell T.\ Hansen, Ben Hörz, and Daniel Mohler for
useful discussions and for some independent checks of our methods. Our
calculations were performed on the HPC cluster ``Wilson'' at the
Institute for Nuclear Physics, University of Mainz, the cluster
``Clover'' on the Helmholtz Institute Mainz, and on the BG/Q computer
JUQUEEN~\cite{juqueen} at JSC, J\"ulich. The authors gratefully
acknowledge the support of the John von Neumann Institute for
Computing and Gauss Centre for Supercomputing
e.V.\ (\url{http://www.gauss-centre.eu}) for project HMZ21. This work
was supported by Deutsche Forschungsgemeinschaft (SFB\,443 and
SFB\,1044) and the Rhineland-Palatinate Research
Initiative. P.M.J. acknowledges support from the Department of
Theoretical Physics (DTP), TIFR. T.D.R. was supported by DFG Grant
No. HA4470/3-1. We are grateful to our colleagues within the CLS
initiative for sharing ensembles.
\end{acknowledgments}

\newpage

\appendix
\section{Fit results \label{sec:appendix}}

In Tables\,\ref{lambda} and \ref{octet} we provide our mass estimates
for single baryons in the SU(3)-symmetric and broken cases,
respectively. Dibaryon energy levels, as determined from fits to the
effective energy, are listed in Tables \ref{singlet} and
\ref{27plet}. The corresponding energy difference with the
$\Lambda\Lambda$ threshold is shown in Tables \ref{singlet_diff} and
\ref{27plet_diff}.

\begingroup
\renewcommand*{\arraystretch}{1.7}
\begin{table}[h]
\centering
\caption{\label{lambda} Mass estimates of the $\Lambda$ hyperon (in
  lattice units) in the SU(3)-symmetric case. The rightmost column
  shows the fit range. Ensembles marked by an asterisk represent
  results determined distillation and LapH smearing.}
\begin{tabular}{D{.}{}{0.2} | c c }
\hline\hline
\multicolumn{1}{c|}{Ensemble} & $am_\Lambda$ & Fit range  \\
\hline \hline
.\text{A1}     & 0.6560(23) & [16,25] \\
\hline
.\text{E1}     & 0.6763(09) & [15,20] \\
.\text{E1}^*   & 0.6751(09) & [12,20] \\
\hline
.\text{N1}     & 0.4538(20) & [25,38] \\
\hline\hline
\end{tabular}
\end{table}
\endgroup

\begin{figure*}
\leavevmode
\includegraphics[width=0.49 \textwidth]{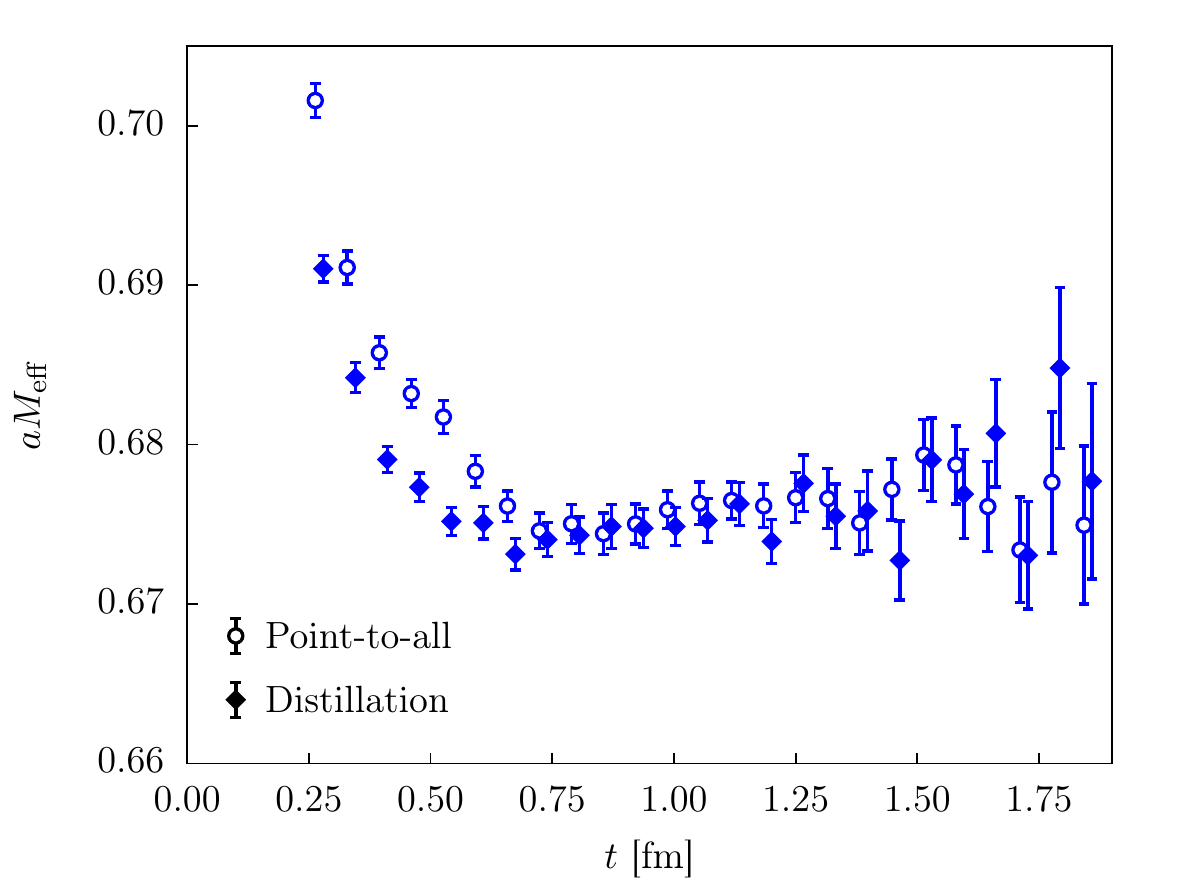}
\hfill
\includegraphics[width=0.49 \textwidth]{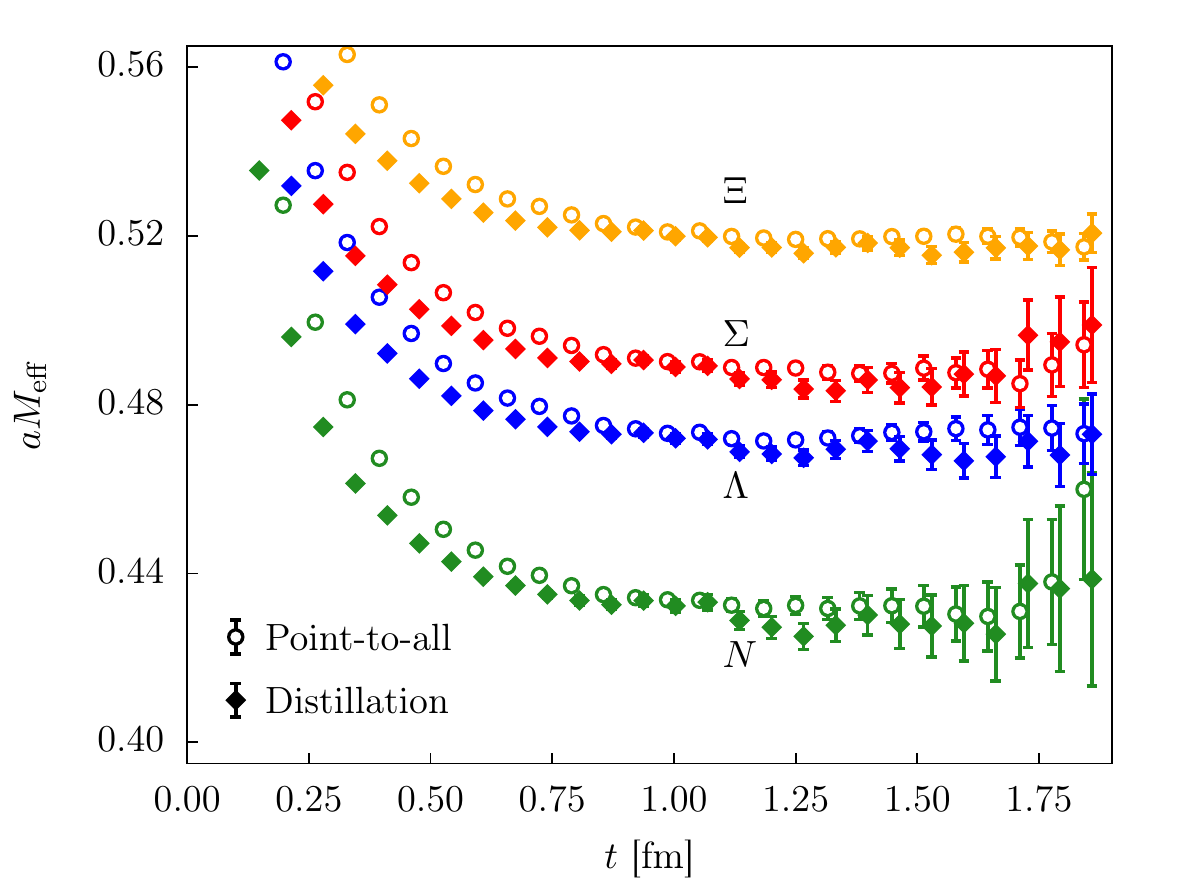}
\caption{\label{fig:octet} Left: Effective masses for the $\Lambda$
  hyperon in the SU(3)-symmetric case (ensemble E1). Right: effective
  masses for octet baryons, i.e. the nucleon, $\Lambda$, $\Sigma$ and
  $\Xi$ in the SU(3)-broken case (ensemble E5). Diamonds and open
  circles denote results obtained using distillation and point
  sources, respectively. For the latter, the narrow smearing width was
  used (see Section~\ref{sec:interpolators}).}
\end{figure*}

The fitted mass estimates for the octet baryons, obtained using either
point sources or distillation, differ at the level of about two
standard deviations (see Table~\ref{octet}). As can be inferred easily
from Fig.~\ref{fig:octet}, the effective masses obtained using either
type of source agree within errors, and hence the differences observed
among the estimates in Table~\ref{octet} are commensurate with
statistical fluctuations.

\begingroup
\renewcommand*{\arraystretch}{1.9}
\begin{table}[h]
  \centering
  \caption{\label{octet} Mass estimates of octet baryons on ensemble
    E5, which corresponds to the SU(3)-broken situation. We list the
    results obtained using point sources and distillation, the latter
    being indicated by an asterisk. The estimates are in lattice units
    along with the relevant fit ranges.}
  \begin{tabular}{c | c c | c c }
    \hline\hline
     & E5 & Fit range & E5$^*$ & Fit range \\
    \hline \hline
    $am_N$        & 0.4330(11) & [15,25] & 0.4302(15) & [15,21] \\
    \hline
    $am_\Lambda$  & 0.4727(07) & [15,25] & 0.4701(10) & [15,21] \\
    \hline
    $am_\Sigma$   & 0.4893(08) & [15,25] & 0.4870(11) & [15,21] \\
    \hline
    $am_\Xi$      & 0.5201(05) & [15,25] & 0.5181(07) & [15,21] \\
    \hline\hline
    \end{tabular}
\end{table}
\endgroup

\begingroup
\renewcommand*{\arraystretch}{1.7}
\begin{table*}[t]
\centering
\caption{\label{singlet} Fitted energies of the ground state for the
  singlet. Results determined using distillation are listed in rows
  labeled E$1^*$ and E$5^*$.}
\begin{tabular}{D{.}{}{0.2} | D{.}{.}{2.7} c | D{.}{.}{2.7} c | D{.}{.}{2.7} c | D{.}{.}{2.7} c}
\hline
\hline
\multicolumn{1}{c|}{\multirow{2}{*}{Ensemble}} & 
\multicolumn{2}{c|}{$\vec{D}^2=0$} &  
\multicolumn{2}{c|}{$\vec{D}^2=1$} &
\multicolumn{2}{c|}{$\vec{D}^2=2$} &
\multicolumn{2}{c}{$\vec{D}^2=3$} \\
& \multicolumn{1}{c}{$aE_{\text{singlet}}$} & Fit range
& \multicolumn{1}{c}{$aE_{\text{singlet}}$} & Fit range
& \multicolumn{1}{c}{$aE_{\text{singlet}}$} & Fit range
& \multicolumn{1}{c}{$aE_{\text{singlet}}$} & Fit range
\\
\hline\hline
.\text{E1}   & 1.3358(66) & [14,21] & 1.3562(58) & [13,19] & 1.3763(60) & [12,19] & 1.3985(74) & [12,19] \\
.\text{E1}^* & 1.3371(18) & [12,20] &\multicolumn{1}{c}{-}&-&\multicolumn{1}{c}{-}&-&\multicolumn{1}{c}{-}&- \\
\hline
.\text{A1}   & 1.279(20)  & [11,16] & 1.303(20)  & [11,17] & 1.321(21)  & [11,15] & 1.341(23)  & [11,15] \\
\hline
.\text{N1}   & 0.888(10)  & [20,27] & 0.904(11)  & [20,27] & 0.9170(77) & [18,23] & 0.9341(53) & [16,21] \\
\hline
.\text{E5}   & 0.9363(70) & [15,19] & 0.9659(68) & [15,19] & 0.974(16)  & [17,20] & 0.985(20)  & [17,21] \\
.\text{E5}^* & 0.9340(26) & [15,20] &\multicolumn{1}{c}{-}&-&\multicolumn{1}{c}{-}&-&\multicolumn{1}{c}{-}&- \\
\hline\hline
\end{tabular}
\end{table*}
\endgroup

\begingroup
\renewcommand*{\arraystretch}{1.7}
\begin{table*}[t]
\centering
\caption{\label{singlet_diff} Ground-state energy differences
  $a(2m_\Lambda - E_\text{cm})$ for the singlet. Results determined using
  distillation are listed in rows labeled E$1^*$ and E$5^*$. The
  rightmost column contains the results for the binding energy
  determined via L\"uscher's finite-volume formalism.}
\begin{tabular}{D{.}{}{0.2} | D{.}{.}{3.9} | D{.}{.}{3.9} | D{.}{.}{3.9} | D{.}{.}{3.9} | c}
\hline
\hline
\multicolumn{1}{c|}{Ensemble} &
\multicolumn{1}{c|}{$\vec{D}^2=0$} &  
\multicolumn{1}{c|}{$\vec{D}^2=1$} &
\multicolumn{1}{c|}{$\vec{D}^2=2$} &
\multicolumn{1}{c|}{$\vec{D}^2=3$} &
FV analysis \\
\hline\hline
.\text{E1}   & 0.0168(66) & 0.0107(60)  & 0.0047(62)  & -0.0039(77) & \multirow{2}{*}{0.0062(34)} \\
.\text{E1}^* & 0.0130(7) &\multicolumn{1}{c|}{-}&\multicolumn{1}{c|}{-}&\multicolumn{1}{c|}{-}& \\
\hline
.\text{A1}   & 0.033(20)  & 0.024(21)   & 0.020(22)   & 0.015(24)   & 0.028(25)\hphantom{0} \\
\hline
.\text{N1}   & 0.019(10)  & 0.014(11)   & 0.009(7)    & 0.001(5)    & 0.016(13)\hphantom{0} \\
\hline
.\text{E5}   & 0.0091(69) & -0.0004(68) & 0.0118(164) & 0.0214(214) & \multirow{2}{*}{-} \\
.\text{E5}^* & 0.0063(18) &\multicolumn{1}{c|}{-}&\multicolumn{1}{c|}{-}&\multicolumn{1}{c|}{-}& \\
\hline\hline
\end{tabular}
\end{table*}
\endgroup

\begingroup
\renewcommand*{\arraystretch}{1.7}
\begin{table*}[t]
\centering
\caption{\label{27plet} Fitted energies of the ground state for the
  27-plet. Results determined using distillation are listed in the row
  labeled E$1^*$.}
\begin{tabular}{D{.}{}{0.2} | D{.}{.}{2.7} c | D{.}{.}{2.7} c | D{.}{.}{2.7} c | D{.}{.}{2.7} c}
\hline
\hline
\multicolumn{1}{c|}{\multirow{2}{*}{Ensemble}} & 
\multicolumn{2}{c|}{$\vec{D}^2=0$} &  
\multicolumn{2}{c|}{$\vec{D}^2=1$} &
\multicolumn{2}{c|}{$\vec{D}^2=2$} &
\multicolumn{2}{c}{$\vec{D}^2=3$} \\
& \multicolumn{1}{c}{$aE_{\text{27-plet}}$} & Fit range
& \multicolumn{1}{c}{$aE_{\text{27-plet}}$} & Fit range
& \multicolumn{1}{c}{$aE_{\text{27-plet}}$} & Fit range
& \multicolumn{1}{c}{$aE_{\text{27-plet}}$} & Fit range
\\
\hline\hline
.\text{E1}   & 1.3494(58) & [14,20] & 1.3761(54) & [13,19] & 1.4011(51) & [12,19] & 1.4265(61) & [12,16] \\
.\text{E1}^* & 1.3488(19) & [12,17] &\multicolumn{1}{c}{-}&-&\multicolumn{1}{c}{-}&-&\multicolumn{1}{c}{-}&- \\
\hline
.\text{A1}   & 1.318(11)  & [12,18] & 1.346(12)  & [12,18] & 1.367(12)  & [10,16] & 1.394(15)  & [10,16] \\
\hline
.\text{N1}   & 0.9058(68) & [21,29] & 0.9205(83) & [22,29] & 0.9366(93) & [22,28] & 0.953(11)  & [22,29] \\
\hline
.\text{E5}   & 0.9412(48) & [16,20] & 0.9843(48) & [16,20] & 1.0299(66) & [16,20] & 1.0789(94) & [16,20] \\
.\text{E5}^* & 0.9380(25) & [15,20] &\multicolumn{1}{c}{-}&-&\multicolumn{1}{c}{-}&-&\multicolumn{1}{c}{-}&- \\
\hline\hline
\end{tabular}
\end{table*}
\endgroup

\begingroup
\renewcommand*{\arraystretch}{1.7}
\begin{table*}[t]
\centering
\caption{\label{27plet_diff} Ground-state energy differences
  $a(2m_\Lambda - E_\text{cm})$ for the 27-plet. Results determined using
  distillation are listed in the row labeled E$1^*$.}
\begin{tabular}{D{.}{}{0.2} | D{.}{.}{3.9} | D{.}{.}{3.9} | D{.}{.}{3.9} | D{.}{.}{3.9}}
\hline
\hline
\multicolumn{1}{c|}{Ensemble} &
\multicolumn{1}{c|}{$\vec{D}^2=0$} &  
\multicolumn{1}{c|}{$\vec{D}^2=1$} &
\multicolumn{1}{c|}{$\vec{D}^2=2$} &
\multicolumn{1}{c}{$\vec{D}^2=3$} \\
\hline\hline
.\text{E1}   & 0.0032(56) & -0.0094(53) & -0.0207(51) & -0.0328(63) \\
.\text{E1}^* & 0.0014(8) &\multicolumn{1}{c|}{-}&\multicolumn{1}{c|}{-}&\multicolumn{1}{c}{-} \\
\hline
.\text{A1}   & -0.006(11) & -0.019(12)  & -0.027(13)  & -0.040(16) \\
\hline
.\text{N1}   & 0.002(6)   & -0.004(7)   & -0.011(9)   & -0.018(10) \\
\hline
.\text{E5}   & 0.0042(46) & -0.0191(47) & -0.0464(67) & -0.0785(98) \\
.\text{E5}^* & 0.0023(17) &\multicolumn{1}{c|}{-}&\multicolumn{1}{c|}{-}&\multicolumn{1}{c}{-} \\
\hline\hline
\end{tabular}
\end{table*}
\endgroup

\newpage

\begin{thebibliography}{10}

\bibitem{Lebed:2016hpi}
R.~F. Lebed, R.~E. Mitchell, and E.~S. Swanson, ``{Heavy-Quark QCD Exotica},''
  \href{http://dx.doi.org/10.1016/j.ppnp.2016.11.003}{Prog. Part. Nucl. Phys.
  {\bfseries 93} (2017) 143--194},
\href{http://arxiv.org/abs/1610.04528}{{\ttfamily arXiv:1610.04528 [hep-ph]}}.

\bibitem{Aaij:2015tga}
{\bfseries LHCb} Collaboration, R.~Aaij {\em et~al.}, ``{Observation of $J/\psi
  p$ Resonances Consistent with Pentaquark States in $\Lambda_b^0 \to J/\psi
  K^- p$ Decays},''
  \href{http://dx.doi.org/10.1103/PhysRevLett.115.072001}{Phys. Rev. Lett.
  {\bfseries 115} (2015) 072001},
\href{http://arxiv.org/abs/1507.03414}{{\ttfamily arXiv:1507.03414 [hep-ex]}}.

\bibitem{Jaffe:1976yi}
R.~L. Jaffe, ``Perhaps a stable dihyperon,''
  \href{http://dx.doi.org/10.1103/PhysRevLett.38.195}{Phys. Rev. Lett.
  {\bfseries 38} (1977) 195--198}.
[Erratum: Phys. Rev. Lett. 38, 617 (1977)].

\bibitem{Takahashi:2001nm}
H.~Takahashi {\em et~al.}, ``{Observation of a
  $\isotope[6][\Lambda\Lambda]{He}$ double hypernucleus},''
\href{http://dx.doi.org/10.1103/PhysRevLett.87.212502}{Phys. Rev. Lett.
  {\bfseries 87} (2001) 212502}.

\bibitem{Nakazawa:2010zza}
{\bfseries KEK-E176, E373, and J-PARC-E07} Collaboration, K.~Nakazawa,
  ``{Double-$\Lambda$ hypernuclei via the $\Xi^-$ hyperon capture at rest
  reaction in a hybrid emulsion},''
\href{http://dx.doi.org/10.1016/j.nuclphysa.2010.01.195}{Nucl. Phys. A
  {\bfseries 835} (2010) 207--214}.

\bibitem{Kim:2013vym}
{\bfseries Belle} Collaboration, B.~H. Kim {\em et~al.}, ``{Search for an
  $H$-dibaryon with mass near $2m_\Lambda$ in $\Upsilon(1S)$ and $\Upsilon(2S)$
  decays},'' \href{http://dx.doi.org/10.1103/PhysRevLett.110.222002}{Phys. Rev.
  Lett. {\bfseries 110} (2013) 222002},
\href{http://arxiv.org/abs/1302.4028}{{\ttfamily arXiv:1302.4028 [hep-ex]}}.

\bibitem{Mackenzie:1985vv}
P.~B. Mackenzie and H.~B. Thacker, ``Evidence against a stable dibaryon from
  lattice {QCD},''
\href{http://dx.doi.org/10.1103/PhysRevLett.55.2539}{Phys. Rev. Lett.
  {\bfseries 55} (1985) 2539}.

\bibitem{Iwasaki:1987db}
Y.~Iwasaki, T.~Yoshie, and Y.~Tsuboi, ``The {$H$} dibaryon in lattice {QCD},''
\href{http://dx.doi.org/10.1103/PhysRevLett.60.1371}{Phys. Rev. Lett.
  {\bfseries 60} (1988) 1371--1374}.

\bibitem{Luo:2007zzb}
Z.-H. Luo, M.~Loan, and X.-Q. Luo, ``{H-dibaryon from lattice QCD with improved
  anisotropic actions},''
  \href{http://dx.doi.org/10.1142/S0217732307023171}{Mod. Phys. Lett.
  {\bfseries A22} (2007) 591--597},
\href{http://arxiv.org/abs/0803.3171}{{\ttfamily arXiv:0803.3171 [hep-lat]}}.

\bibitem{Luo:2011ar}
Z.-H. Luo, M.~Loan, and Y.~Liu, ``{Search for the $H$ dibaryon on the
  lattice},'' \href{http://dx.doi.org/10.1103/PhysRevD.84.034502}{Phys. Rev. D
  {\bfseries 84} (2011) 034502},
\href{http://arxiv.org/abs/1106.1945}{{\ttfamily arXiv:1106.1945 [hep-lat]}}.

\bibitem{Pochinsky:1998zi}
A.~Pochinsky, J.~W. Negele, and B.~Scarlet, ``{Lattice study of the $H$
  dibaryon},'' \href{http://dx.doi.org/10.1016/S0920-5632(99)85040-3}{Nucl.
  Phys. Proc. Suppl. {\bfseries 73} (1999) 255--257},
\href{http://arxiv.org/abs/hep-lat/9809077}{{\ttfamily arXiv:hep-lat/9809077
  [hep-lat]}}.

\bibitem{Wetzorke:1999rt}
I.~Wetzorke, F.~Karsch, and E.~Laermann, ``{Further evidence for an unstable H
  dibaryon?},'' \href{http://dx.doi.org/10.1016/S0920-5632(00)91628-1}{Nucl.
  Phys. Proc. Suppl. {\bfseries 83} (2000) 218--220},
\href{http://arxiv.org/abs/hep-lat/9909037}{{\ttfamily arXiv:hep-lat/9909037
  [hep-lat]}}.

\bibitem{Wetzorke:2002mx}
I.~Wetzorke and F.~Karsch, ``{The H dibaryon on the lattice},''
  \href{http://dx.doi.org/10.1016/S0920-5632(03)01531-7}{Nucl. Phys. Proc.
  Suppl. {\bfseries 119} (2003) 278--280},
\href{http://arxiv.org/abs/hep-lat/0208029}{{\ttfamily arXiv:hep-lat/0208029
  [hep-lat]}}.

\bibitem{Beane:2011zpa}
S.~R. Beane {\em et~al.}, ``Present constraints on the {H}-dibaryon at the
  physical point from lattice {QCD},''
  \href{http://dx.doi.org/10.1142/S0217732311036978}{Mod. Phys. Lett. A
  {\bfseries 26} (2011) 2587--2595},
\href{http://arxiv.org/abs/1103.2821}{{\ttfamily arXiv:1103.2821 [hep-lat]}}.

\bibitem{Francis:2013lva}
A.~Francis, C.~Miao, T.~D. Rae, and H.~Wittig, ``Two-baryon correlation
  functions in 2-flavour {QCD},''
  \href{http://dx.doi.org/10.22323/1.187.0440}{PoS {\bfseries LATTICE2013}
  (2014) 440},
\href{http://arxiv.org/abs/1311.3933}{{\ttfamily arXiv:1311.3933 [hep-lat]}}.

\bibitem{Green:2014dea}
J.~Green, A.~Francis, P.~Junnarkar, C.~Miao, T.~Rae, and H.~Wittig, ``{Search
  for a bound H-dibaryon using local six-quark interpolating operators},''
  \href{http://dx.doi.org/10.22323/1.214.0107}{PoS {\bfseries LATTICE2014}
  (2014) 107},
\href{http://arxiv.org/abs/1411.1643}{{\ttfamily arXiv:1411.1643 [hep-lat]}}.

\bibitem{Junnarkar:2015jyf}
P.~Junnarkar, A.~Francis, J.~Green, C.~Miao, T.~Rae, and H.~Wittig, ``Search
  for the {H}-dibaryon in two flavor lattice {QCD},''
  \href{http://dx.doi.org/10.22323/1.253.0079}{PoS {\bfseries CD15} (2015)
  079}, \href{http://arxiv.org/abs/1511.01849}{{\ttfamily arXiv:1511.01849
  [hep-lat]}}.
[PoS LATTICE2015, 082 (2016)].

\bibitem{Beane:2010hg}
{\bfseries NPLQCD} Collaboration, S.~R. Beane {\em et~al.}, ``Evidence for a
  bound {$H$} dibaryon from lattice {QCD},''
  \href{http://dx.doi.org/10.1103/PhysRevLett.106.162001}{Phys. Rev. Lett.
  {\bfseries 106} (2011) 162001},
\href{http://arxiv.org/abs/1012.3812}{{\ttfamily arXiv:1012.3812 [hep-lat]}}.

\bibitem{Beane:2011iw}
{\bfseries NPLQCD} Collaboration, S.~R. Beane, E.~Chang, W.~Detmold, H.~W. Lin,
  T.~C. Luu, K.~Orginos, A.~Parreño, M.~J. Savage, A.~Torok, and
  A.~Walker-Loud, ``The deuteron and exotic two-body bound states from lattice
  {QCD},'' \href{http://dx.doi.org/10.1103/PhysRevD.85.054511}{Phys. Rev. D
  {\bfseries 85} (2012) 054511},
\href{http://arxiv.org/abs/1109.2889}{{\ttfamily arXiv:1109.2889 [hep-lat]}}.

\bibitem{Beane:2012vq}
{\bfseries NPLQCD} Collaboration, S.~R. Beane, E.~Chang, S.~D. Cohen,
  W.~Detmold, H.~W. Lin, T.~C. Luu, K.~Orginos, A.~Parreño, M.~J. Savage, and
  A.~Walker-Loud, ``Light nuclei and hypernuclei from quantum chromodynamics in
  the limit of {SU(3)} flavor symmetry,''
  \href{http://dx.doi.org/10.1103/PhysRevD.87.034506}{Phys. Rev. D {\bfseries
  87} (2013) 034506},
\href{http://arxiv.org/abs/1206.5219}{{\ttfamily arXiv:1206.5219 [hep-lat]}}.

\bibitem{Inoue:2010hs}
{\bfseries HAL QCD} Collaboration, T.~Inoue, N.~Ishii, S.~Aoki, T.~Doi,
  T.~Hatsuda, Y.~Ikeda, K.~Murano, H.~Nemura, and K.~Sasaki, ``Baryon-baryon
  interactions in the flavor {SU(3)} limit from full {QCD} simulations on the
  lattice,'' \href{http://dx.doi.org/10.1143/PTP.124.591}{Prog. Theor. Phys.
  {\bfseries 124} (2010) 591--603},
\href{http://arxiv.org/abs/1007.3559}{{\ttfamily arXiv:1007.3559 [hep-lat]}}.

\bibitem{Inoue:2010es}
{\bfseries HAL QCD} Collaboration, T.~Inoue, N.~Ishii, S.~Aoki, T.~Doi,
  T.~Hatsuda, Y.~Ikeda, K.~Murano, H.~Nemura, and K.~Sasaki, ``Bound {$H$}
  dibaryon in flavor {SU(3)} limit of lattice {QCD},''
  \href{http://dx.doi.org/10.1103/PhysRevLett.106.162002}{Phys. Rev. Lett.
  {\bfseries 106} (2011) 162002},
\href{http://arxiv.org/abs/1012.5928}{{\ttfamily arXiv:1012.5928 [hep-lat]}}.

\bibitem{Inoue:2011ai}
{\bfseries HAL QCD} Collaboration, T.~Inoue, S.~Aoki, T.~Doi, T.~Hatsuda,
  Y.~Ikeda, N.~Ishii, K.~Murano, H.~Nemura, and K.~Sasaki, ``Two-baryon
  potentials and {$H$}-dibaryon from 3-flavor lattice {QCD} simulations,''
  \href{http://dx.doi.org/10.1016/j.nuclphysa.2012.02.008}{Nucl. Phys. A
  {\bfseries 881} (2012) 28--43},
\href{http://arxiv.org/abs/1112.5926}{{\ttfamily arXiv:1112.5926 [hep-lat]}}.

\bibitem{Sasaki:2016gpc}
K.~Sasaki {\em et~al.}, ``{First results of baryon interactions from lattice
  QCD with physical masses (3) -- Strangeness $S=-2$ two-baryon system},''
\href{http://dx.doi.org/10.22323/1.251.0088}{PoS {\bfseries LATTICE2015} (2016)
  088}.

\bibitem{Sasaki:2018mzh}
K.~Sasaki, S.~Aoki, T.~Doi, S.~Gongyo, T.~Hatsuda, Y.~Ikeda, T.~Inoue,
  T.~Iritani, N.~Ishii, and T.~Miyamoto, ``{Lattice QCD studies on baryon
  interactions in the strangeness -2 sector with physical quark masses},''
\href{http://dx.doi.org/10.1051/epjconf/201817505010}{EPJ Web Conf. {\bfseries
  175} (2018) 05010}.

\bibitem{Haidenbauer:2011ah}
J.~Haidenbauer and U.-G. Meissner, ``{To bind or not to bind: The H-dibaryon in
  light of chiral effective field theory},''
  \href{http://dx.doi.org/10.1016/j.physletb.2011.10.070}{Phys. Lett.
  {\bfseries B706} (2011) 100--105},
\href{http://arxiv.org/abs/1109.3590}{{\ttfamily arXiv:1109.3590 [hep-ph]}}.

\bibitem{Haidenbauer:2011za}
J.~Haidenbauer and U.~G. Meissner, ``{Exotic bound states of two baryons in
  light of chiral effective field theory},''
  \href{http://dx.doi.org/10.1016/j.nuclphysa.2012.01.021}{Nucl. Phys.
  {\bfseries A881} (2012) 44--61},
\href{http://arxiv.org/abs/1111.4069}{{\ttfamily arXiv:1111.4069 [nucl-th]}}.

\bibitem{Michael:1985ne}
C.~Michael, ``{Adjoint Sources in Lattice Gauge Theory},''
\href{http://dx.doi.org/10.1016/0550-3213(85)90297-4}{Nucl. Phys. {\bfseries
  B259} (1985) 58--76}.

\bibitem{Luscher:1990ck}
M.~L{\"u}scher and U.~Wolff, ``How to calculate the elastic scattering matrix
  in two-dimensional quantum field theories by numerical simulation,''
\href{http://dx.doi.org/10.1016/0550-3213(90)90540-T}{Nucl. Phys. B {\bfseries
  339} (1990) 222--252}.

\bibitem{Blossier:2009kd}
B.~Blossier, M.~Della~Morte, G.~von Hippel, T.~Mendes, and R.~Sommer, ``{On the
  generalized eigenvalue method for energies and matrix elements in lattice
  field theory},''
\href{http://dx.doi.org/10.1088/1126-6708/2009/04/094}{JHEP {\bfseries 04}
  (2009) 094}.

\bibitem{Peardon:2009gh}
{\bfseries Hadron Spectrum} Collaboration, M.~Peardon, J.~Bulava, J.~Foley,
  C.~Morningstar, J.~Dudek, R.~G. Edwards, B.~Joo, H.-W. Lin, D.~G. Richards,
  and K.~J. Juge, ``{A Novel quark-field creation operator construction for
  hadronic physics in lattice QCD},''
  \href{http://dx.doi.org/10.1103/PhysRevD.80.054506}{Phys. Rev. D {\bfseries
  80} (2009) 054506},
\href{http://arxiv.org/abs/0905.2160}{{\ttfamily arXiv:0905.2160 [hep-lat]}}.

\bibitem{Aoki:2016frl}
S.~Aoki {\em et~al.}, ``{Review of lattice results concerning low-energy
  particle physics},''
  \href{http://dx.doi.org/10.1140/epjc/s10052-016-4509-7}{Eur. Phys. J.
  {\bfseries C77} no.~2, (2017) 112},
\href{http://arxiv.org/abs/1607.00299}{{\ttfamily arXiv:1607.00299 [hep-lat]}}.

\bibitem{Sheikholeslami:1985ij}
B.~Sheikholeslami and R.~Wohlert, ``{Improved Continuum Limit Lattice Action
  for QCD with Wilson Fermions},''
\href{http://dx.doi.org/10.1016/0550-3213(85)90002-1}{Nucl. Phys. {\bfseries
  B259} (1985) 572}.

\bibitem{Luscher:2005rx}
M.~L{\"u}scher, ``{Schwarz-preconditioned HMC algorithm for two-flavour lattice
  QCD},'' \href{http://dx.doi.org/10.1016/j.cpc.2004.10.004}{Comput. Phys.
  Commun. {\bfseries 165} (2005) 199--220},
\href{http://arxiv.org/abs/hep-lat/0409106}{{\ttfamily arXiv:hep-lat/0409106
  [hep-lat]}}.

\bibitem{Luscher:2007es}
M.~L{\"u}scher, ``{Deflation acceleration of lattice QCD simulations},''
  \href{http://dx.doi.org/10.1088/1126-6708/2007/12/011}{JHEP {\bfseries 12}
  (2007) 011},
\href{http://arxiv.org/abs/0710.5417}{{\ttfamily arXiv:0710.5417 [hep-lat]}}.

\bibitem{Marinkovic:2010eg}
M.~Marinkovic and S.~Schaefer, ``{Comparison of the mass preconditioned HMC and
  the DD-HMC algorithm for two-flavour QCD},''
  \href{http://dx.doi.org/10.22323/1.105.0031}{PoS {\bfseries LATTICE2010}
  (2010) 031},
\href{http://arxiv.org/abs/1011.0911}{{\ttfamily arXiv:1011.0911 [hep-lat]}}.

\bibitem{Jansen:1998mx}
{\bfseries ALPHA} Collaboration, K.~Jansen and R.~Sommer, ``{O($a$) improvement
  of lattice QCD with two flavors of Wilson quarks},''
  \href{http://dx.doi.org/10.1016/S0550-3213(98)00396-4}{Nucl. Phys. {\bfseries
  B530} (1998) 185--203},
  \href{http://arxiv.org/abs/hep-lat/9803017}{{\ttfamily arXiv:hep-lat/9803017
  [hep-lat]}}.
[Erratum: Nucl. Phys. B643 (2002) 517].

\bibitem{Fritzsch:2012wq}
P.~Fritzsch, F.~Knechtli, B.~Leder, M.~Marinkovic, S.~Schaefer, R.~Sommer, and
  F.~Virotta, ``{The strange quark mass and Lambda parameter of two flavor
  QCD},'' \href{http://dx.doi.org/10.1016/j.nuclphysb.2012.07.026}{Nucl. Phys.
  {\bfseries B865} (2012) 397--429},
\href{http://arxiv.org/abs/1205.5380}{{\ttfamily arXiv:1205.5380 [hep-lat]}}.

\bibitem{OpenQCD}
M.~Lüscher and S.~Schaefer, ``{OpenQCD}.''
\newblock \url{http://luscher.web.cern.ch/luscher/openQCD/}.

\bibitem{PRIMME}
A.~Stathopoulos and J.~R. McCombs, ``{PRIMME}: {PR}econditioned {I}terative
  {M}ulti{M}ethod {E}igensolver: Methods and software description,''
  \href{http://dx.doi.org/10.1145/1731022.1731031}{ACM Transactions on
  Mathematical Software {\bfseries 37} no.~2, (2010) 21:1--21:30}.

\bibitem{Edwards:2004sx}
{\bfseries SciDAC, LHPC, UKQCD} Collaboration, R.~G. Edwards and B.~Joó,
  ``{The Chroma software system for lattice QCD},''
  \href{http://dx.doi.org/10.1016/j.nuclphysbps.2004.11.254}{Nucl. Phys. Proc.
  Suppl. {\bfseries 140} (2005) 832},
\href{http://arxiv.org/abs/hep-lat/0409003}{{\ttfamily arXiv:hep-lat/0409003
  [hep-lat]}}.

\bibitem{Blum:2012uh}
T.~Blum, T.~Izubuchi, and E.~Shintani, ``{New class of variance-reduction
  techniques using lattice symmetries},''
  \href{http://dx.doi.org/10.1103/PhysRevD.88.094503}{Phys. Rev. D {\bfseries
  88} (2013) 094503},
\href{http://arxiv.org/abs/1208.4349}{{\ttfamily arXiv:1208.4349 [hep-lat]}}.

\bibitem{Donoghue:1986zd}
E.~Golowich, J.~Donoghue, and B.~R. Holstein, ``{Weak Decays of the $H$
  Dibaryon},''
\href{http://dx.doi.org/10.1103/PhysRevD.34.3434}{Phys. Rev. D {\bfseries 34}
  (1986) 3434}.

\bibitem{Golowich:1992zw}
E.~Golowich and T.~Sotirelis, ``{O($\alpha_s^2$) mass contributions to the $H$
  dibaryon in a truncated bag model},''
\href{http://dx.doi.org/10.1103/PhysRevD.46.354}{Phys. Rev. D {\bfseries 46}
  (1992) 354--363}.

\bibitem{Jaffe:2004ph}
R.~L. Jaffe, ``{Exotica},''
  \href{http://dx.doi.org/10.1016/j.physrep.2004.11.005}{Phys. Rept. {\bfseries
  409} (2005) 1--45},
\href{http://arxiv.org/abs/hep-ph/0409065}{{\ttfamily arXiv:hep-ph/0409065
  [hep-ph]}}.

\bibitem{Gusken:1989qx}
S.~G{\"u}sken, ``{A Study of smearing techniques for hadron
  correlationfunctions},''
\href{http://dx.doi.org/10.1016/0920-5632(90)90273-W}{Nucl. Phys. Proc. Suppl.
  {\bfseries 17} (1990) 361--364}.

\bibitem{Albanese:1987ds}
{\bfseries APE} Collaboration, M.~Albanese {\em et~al.}, ``{Glueball Masses and
  String Tension in Lattice QCD},''
\href{http://dx.doi.org/10.1016/0370-2693(87)91160-9}{Phys. Lett. {\bfseries
  B192} (1987) 163--169}.

\bibitem{Morningstar:2003gk}
C.~Morningstar and M.~J. Peardon, ``{Analytic smearing of SU(3) link variables
  in lattice QCD},'' \href{http://dx.doi.org/10.1103/PhysRevD.69.054501}{Phys.
  Rev. D {\bfseries 69} (2004) 054501},
\href{http://arxiv.org/abs/hep-lat/0311018}{{\ttfamily arXiv:hep-lat/0311018}}.

\bibitem{Hanlon:2018yfv}
A.~Hanlon, A.~Francis, J.~Green, P.~Junnarkar, and H.~Wittig, ``{The $H$
  dibaryon from lattice QCD with SU(3) flavor symmetry},'' in {\em {36th
  International Symposium on Lattice Field Theory (Lattice 2018) East Lansing,
  MI, United States, July 22-28, 2018}}.
\newblock 2018.
\newblock
\href{http://arxiv.org/abs/1810.13282}{{\ttfamily arXiv:1810.13282 [hep-lat]}}.
\newblock

\bibitem{Luscher:1990ux}
M.~L{\"u}scher, ``{Two-particle states on a torus and their relation to the
  scattering matrix},''
\href{http://dx.doi.org/10.1016/0550-3213(91)90366-6}{Nucl. Phys. B {\bfseries
  354} (1991) 531--578}.

\bibitem{Rummukainen:1995vs}
K.~Rummukainen and S.~A. Gottlieb, ``{Resonance scattering phase shifts on a
  non-rest-frame lattice},''
  \href{http://dx.doi.org/10.1016/0550-3213(95)00313-H}{Nucl. Phys. B
  {\bfseries 450} (1995) 397--436},
\href{http://arxiv.org/abs/hep-lat/9503028}{{\ttfamily arXiv:hep-lat/9503028
  [hep-lat]}}.

\bibitem{Gockeler:2012yj}
M.~G{\"o}ckeler, R.~Horsley, M.~Lage, U.-G. Mei{\ss}ner, P.~E.~L. Rakow,
  A.~Rusetsky, G.~Schierholz, and J.~M. Zanotti, ``{Scattering phases for meson
  and baryon resonances on general moving-frame lattices},''
  \href{http://dx.doi.org/10.1103/PhysRevD.86.094513}{Phys. Rev. D {\bfseries
  86} (2012) 094513},
\href{http://arxiv.org/abs/1206.4141}{{\ttfamily arXiv:1206.4141 [hep-lat]}}.

\bibitem{Iritani:2017rlk}
T.~Iritani, S.~Aoki, T.~Doi, T.~Hatsuda, Y.~Ikeda, T.~Inoue, N.~Ishii,
  H.~Nemura, and K.~Sasaki, ``{Are two nucleons bound in lattice QCD for heavy
  quark masses? Consistency check with Lüscher’s finite volume formula},''
  \href{http://dx.doi.org/10.1103/PhysRevD.96.034521}{Phys. Rev. D {\bfseries
  96} (2017) 034521},
\href{http://arxiv.org/abs/1703.07210}{{\ttfamily arXiv:1703.07210 [hep-lat]}}.

\bibitem{Erben:2017hvr}
F.~Erben, J.~Green, D.~Mohler, and H.~Wittig, ``{Towards extracting the
  timelike pion form factor on CLS two-flavour ensembles},''
  \href{http://dx.doi.org/10.1051/epjconf/201817505027}{EPJ Web Conf.
  {\bfseries 175} (2018) 05027},
\href{http://arxiv.org/abs/1710.03529}{{\ttfamily arXiv:1710.03529 [hep-lat]}}.

\bibitem{juqueen}
{Jülich Supercomputing Centre}, ``{JUQUEEN: IBM Blue Gene/Q Supercomputer
  System at the Jülich Supercomputing Centre},''
  \href{http://dx.doi.org/10.17815/jlsrf-1-18}{Journal of large-scale research
  facilities {\bfseries 1} (2015) A1}.

\end{thebibliography}
\providecommand{\href}[2]{#2}\begingroup\raggedright\endgroup

\end{document}